\documentclass[journal]{IEEEtran}
\usepackage{cite} % Tidies up citation numbers.
\usepackage{url} % Provides better formatting of URLs.
\usepackage[utf8]{inputenc} % Allows Turkish characters.
\usepackage{booktabs} % Allows the use of \toprule, \midrule and \bottomrule in tables for horizontal lines
\usepackage{graphicx}

\usepackage{epsfig}
\usepackage{algorithm}
\usepackage{algpseudocode}
\usepackage{amsmath}
\usepackage{amssymb}
\usepackage{url}
\usepackage{caption}
\usepackage{subcaption}
\DeclareCaptionType{copyrightbox}

\algnotext{EndIf}
\algnotext{EndFor}
\algnotext{EndWhile}

\begin{document}
%\begin{titlepage}
\title{MeshCloak: A Map-Based Approach for Personalized Location Privacy}


\author{Hiep H. Nguyen \\
Institute of Research and Development, Duy Tan University\\
P809 7/25 Quang Trung, Danang 550000, Vietnam\\
}
\date{15/5/17}

\maketitle
%\tableofcontents
%\listoffigures
%\listoftables
%\end{titlepage}

\IEEEpeerreviewmaketitle
\begin{abstract}
Protecting location privacy in mobile services has recently received significant consideration as Location-Based Service (LBS) can reveal user locations to attackers. A problem in the existing cloaking schemes is that location vulnerabilities may be exposed when an attacker exploits a street map in their attacks. While both real and synthetic trajectories are based on real street maps, most of previous cloaking schemes assume free space movements to define the distance between users, resulting in the mismatch between privacy models and user movements. In this paper, we present MeshCloak, a novel map-based model for personalized location privacy, which is formulated entirely in map-based setting and resists inference attacks at a minimal performance overhead. The key idea of MeshCloak is to quickly build a sparse constraint graph based on the mutual coverage relationship between queries by pre-computing the distance matrix and applying quadtree search. MeshCloak also takes into account real speed profiles and query frequencies. We evaluate the efficiency and effectiveness of the proposed scheme via a suite of carefully designed experiments on five real maps.

\end{abstract}

\section{Introduction}
\label{sec:intro}
Context awareness in mobile service has been widely adopted in the last decade. A typical example of context-aware mobile service is Location-Based Service (LBS) which exploits geographical positions of mobile users to provide convenient information and guidance (e.g., navigation guidance, point-of-interest queries, traffic alerts, etc.) However, such location-based services raise critical security and privacy issues because attackers can extract a mobile user's personal information based on his/her locations and query contents. To provide location privacy, existing approaches are either \textit{user-centric} (i.e. individual location obfuscation without the anonymizer) or \textit{anonymizer-based} in which an anonymizer cloaks user locations into areas covering multiple users (e.g. at least \textit{k} users in k-anonymity \cite{sweeney2002k}). Regarding the query update frequency in LBS applications, existing algorithms can be classified into two main groups: \textit{sporadic} query (i.e., queries are exposed infrequently and the attacker's goal is to localize users at certain time instants) and \textit{continuous} query (i.e., queries are continuously issued by users and the adversary can track users over time and space). This paper employs the cloaking approach for continuous queries issued at different frequencies.

Personalized location privacy \cite{gedik2008protecting} captures varying location privacy requirements in which each mobile user specifies his anonymity level (\textit{k} value), spatial tolerance, and temporal tolerance. In personalized location privacy, two users $u$ and $v$ are potentially cloaked together if each of them is covered by the spatio-temporal constraint box \cite{gedik2008protecting} of the other user. Such a mutual coverage relationship forms an edge in the constraint graph. Similarly, a group of \textit{k} users are cloaked together if they form a $k$-clique. This strong requirement is called \textit{reciprocity}, first coined in Hilbert Cloak\cite{kalnis2007preventing} (see Section \ref{sec:related}).

In this paper, we present \textit{MeshCloak}, a novel map-based model for personalized location privacy. Compared to CliqueCloak \cite{gedik2008protecting}, MeshCloak differs in two main points. First, while users move on streets, CliqueCloak assumes spatial constraint as a rectangle. This may become unrealistic and an attacker can amount effective inference attacks by applying map-based constraints. MeshCloak fixes this vulnerability by using user speed constraints as spatial tolerance on the street map. Second, CliqueCloak processes incoming queries one-by-one and runs an inefficient search of maximal cliques in the constraint graph. This limits the throughput of the anonymizer. MeshCloak, on the contrary, collects and processes queries in each time unit (e.g. in every second). Using quadtree structure and the fast all-maximal clique listing by Tomita et al. \cite{tomita2006worst}, it reduces the processing time by two orders of magnitude.

\begin{figure}
\centering
\epsfig{file=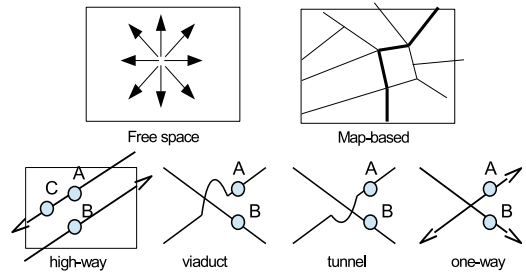, height=1.6in}
\setlength{\abovecaptionskip}{-10pt}
\caption{(Top) free space vs. map-based settings. (Bottom) movement constraints in map-based setting.}
\label{fig:moving-patterns}
\end{figure}

Fig. \ref{fig:moving-patterns} contrasts the movement in free space and in map-based settings. Unlike in free space setting, movements in map-based setting are stipulated by real-world constraints on route direction and speed. The users A and B in Fig. \ref{fig:moving-patterns} are close to each other in free space but this is no longer the case from map-based perspective. Most of previous cloaking schemes omit these shortest path intuitions in their model by allowing user cloaking if the users are within a certain Euclidean distance from each other.

\subsection{Contributions and Scope}

The three main ideas distinguish \textit{MeshCloak} from the previous location privacy schemes. First, the real street map is integrated for searching near-by mobile users to be cloaked together, this idea eliminates the popular and unrealistic assumption of free space movement. Second, a fast distance computation between users is realized via a pre-computed distance matrix and quadtree search. These techniques help to maintain the sparse constraint graph in time nearly linear to the number of users. Third, queries are processed in batches (one batch per second) and maximal cliques in the current time unit are listed quickly by running the Tomita algorithm \cite{tomita2006worst}. 

Compared to prior work, our scheme highlights the following features
\begin{itemize}
\item We formulate the problem of personalized location privacy entirely in map-based setting. To the best of our knowledge, our research is the first to apply movement distance constraint for cloaking purpose. We argue that our problem formulation provides more realistic view on spatial awareness. 
\item We propose a novel cloaking scheme MeshCloak for this problem. MeshCloak incurs only a small processing overhead thanks to the fast computation of the constraint graph and maximal cliques within it. Our scheme can process up to 30,000 queries per second, a throughput much larger than the one reported in \cite{gedik2008protecting} and \cite{pan2012protecting}.
\item We customize the Brinkhoff simulator \cite{brinkhoff2002framework} to support varying query frequencies and realistic user speeds. The performance evaluation on five real maps validates the effectiveness and efficiency of our MeshCloak.
\end{itemize} 

The remainder of the paper is organized as follows. Section \ref{sec:related} briefly reviews related work and limitations. The motivation for the current work is introduced in Section \ref{sec:model}. Section \ref{sec:model} also defines the necessary concepts, mobility model and assumptions in MeshCloak. Section \ref{sec:algo} presents MeshCloak processing steps. Evaluation results are discussed in Section \ref{sec:evaluation}. Section \ref{sec:conclusion} concludes the paper.

\section{Related work}
\label{sec:related}
To motivate our MeshCloak scheme, this section reviews previous LBS privacy models and some limitations.

\subsection{Personalized Location Privacy}
The intuition for LBS privacy of many existing work is ``protection from being brought to the attention of others" by Gavison \cite{gavison1979privacy}, which means safety by \textit{blending yourself into a crowd}. K-anonymity \cite{sweeney2002k} and its extensions like \textit{l-diversity} \cite{machanavajjhala2007diversity}, are therefore broadly investigated in previous location privacy schemes \cite{gruteser2003anonymous,mokbel2006new,kalnis2007preventing,wang2009privacy,wang2012l2p2}. Along with k-anonymity model, to avoid outlier revealing attacks, cloaking areas should satisfy \textit{reciprocity} which is first coined in \textit{Hilbert Cloak}\cite{kalnis2007preventing}. Reciprocity requires that a set of $k$ users must have $\frac{k(k-1)}{2}$ relationships, also called k-clique in graph terms. \textit{Reciprocal framework} \cite{ghinita2010reciprocal} generalized the idea of reciprocity to be adaptable to any existing spatial index on the user locations. Reciprocity appeared earlier under different names of \textit{clique} in \textit{CliqueCloak} \cite{gedik2008protecting} and \textit{k-sharing region} in \cite{chow2007enabling}. For moving objects databases, \cite{abul2008never} defines the Anonymity Set of Trajectories based on the mutual \textit{Co-localization} which also implies reciprocity.

Personalized location k-anonymity brings versatility to users, allowing them to dynamically increase (decrease) privacy requirements when entering (leaving) easily identifiable areas. This concept was first proposed by Gedik and Liu \cite{gedik2008protecting}, employing both \textit{spatial cloaking} (constraint boxes) and \textit{temporal cloaking} (message delaying). The personalized scheme became the \textit{de-facto} and was extended in many ways in LBS privacy protection \cite{chow2007enabling,pan2012protecting} and relational database research \cite{xiao2006personalized}. 

\subsection{Free Space Model}
Location anonymization in free space appears in a large number of existing papers \cite{gruteser2003anonymous}, \cite{gedik2008protecting}, \cite{kalnis2007preventing},   \cite{ghinita2010reciprocal}, \cite{lee2011protecting}, \cite{pan2012protecting}, \cite{amini2011cache}, \cite{wang2012l2p2}. The assumption of free movement in all directions considerably simplifies the spatial privacy model, paving the way for the application of popular tree structures (quad-tree, R-tree) and partitioning techniques (grid-like partitioning, Hilbert filling curves). Neighborhood relationship in this setting is also reduced to simple geometrical operations, e.g. point-in-rectangle checking. This simplification certainly ignores many real-world movement constraints like high-ways, viaducts, tunnels, one-way routes, impasses and so on (Fig. 1).

Moving in constrained space attracts a bit less attention \cite{wang2009privacy}, \cite{ying2014protecting}, \cite{palanisamy2011mobimix}. \textit{XSTAR} \cite{wang2009privacy} and \textit{PLPCA} \cite{ying2014protecting} support sporadic query and use k-anonymity along with \textit{l-diversity} \cite{machanavajjhala2007diversity} of street segments. \textit{MobiMix} \cite{palanisamy2011mobimix} exploits the ``mix zone'' concept to cope with timing tracking attack and may have the problem of space-time intersection rarity.

\subsection{Moving-Together Implication}
In continuous query model, typical attacks like \textit{location-dependent attacks} \cite{cheng2006preserving}, \cite{pan2012protecting} and \textit{query tracking attacks} \cite{chow2007enabling} imply ``moving-together'' requirement. The two solutions \textit{patching} and \textit{delaying} in \cite{cheng2006preserving} induce large cloaking areas over time to keep non-decreasing uncertainty of cloaked location. \textit{ICliqueCloak} \cite{pan2012protecting} suffers from a similar issue in which cloaking areas may grow up to 80\% of the map area. The \textit{memorization} property proposed in \cite{chow2007enabling} is strict enough to prevent query tracking attacks. Cloaking areas reported in \cite{chow2007enabling} are up to 11\% of the map area thanks to the flexible group join/leave operations. However, both \cite{chow2007enabling} and \cite{pan2012protecting} are formulated in free space setting. %To overcome the above limitation, a viable solution should be less conservative, allowing mobile users to change their cloaking set more frequently over time provided that a minimum number of users must go with him into the new cloaking set. Our scheme \textit{MeshCloak} favours this direction.

%\subsection{Caching and Historical Data}
%Caching can help preserving privacy partially \cite{meyerowitz2009hiding} or fully \cite{amini2011cache}. In \textit{CacheCloak} \cite{meyerowitz2009hiding}, mobility prediction is used to \textit{do a prospective form of path confusion}. The anonymity is assured by implicit collaboration among users. The target user cannot be tracked by the LBS server because its newly predicted paths (in the case of cache miss) are connected to unexpired paths of some users (maybe including herself). This is where two problems occur: extrapolating a too long predicted path may leak information about user density and different query types on different paths are not mentioned. Fully caching approach of \textit{Cach\'{e}}\cite{amini2011cache} eliminates the collaborative nature of conventional location privacy models. \textit{Cach\'{e}} only addresses text data due to the constraints of device storage and update bandwidth.
%
%Historical data, especially mobility traces, is often used to model \textit{probabilistic location privacy}, \textit{semantic location privacy} \cite{xu2007location}, \cite{xu2009feeling}, \cite{meyerowitz2009hiding},  \cite{lee2011protecting} and \textit{user mobility profiles} for inference attacks \cite{shokri2011quantifying}, \cite{shokri2012protecting}. However, those approaches need large-scale real and unbiased datasets, e.g. Call Data Records from a nation-wide US cellular provider as in \cite{zang2011anonymization}. %In our model, each user has a mobility pattern formed from his set of simulated historical traces. 

% %
\section{MeshCloak Privacy Model}
\label{sec:model}

\subsection{Motivation}
\begin{figure}
\centering
\epsfig{file=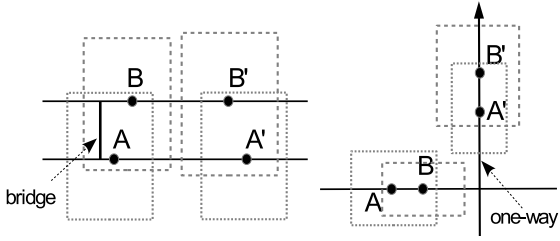, height=1.5in}
\setlength{\abovecaptionskip}{-10pt}
\setlength{\belowcaptionskip}{-10pt}
\caption{Free space vulnerability in CliqueCloak \cite{gedik2008protecting}}
\label{fig:cliquecloak-limit}
\end{figure}

In this section, we explain some limitations of free space assumption in existing schemes. Fig. \ref{fig:cliquecloak-limit} illustrates such problematic cases in CliqueCloak \cite{gedik2008protecting}. At time $t+\Delta t$, users A and B move to new locations A$'$ and B$'$. Using constraint boxes as in CliqueCloak, A$'$ and B$'$ are still considered ``close'' and they form an edge in the constraint graph. However, an attacker by using the map can reveal that the shortest path between A$'$ and B$'$ increases considerably at $t+\Delta t$ and may remove B from cloaking set of A and vice-versa. In the left figure, A$'$ and B$'$ move far away the bridge so the shortest path between them gets longer. In the right figure, the one-way suggests that A$'$ is near B$'$ but not vice-versa. We observe that closeness in terms of shortest path implies closeness in free space but the converse is not true.

\begin{figure}
\centering
\epsfig{file=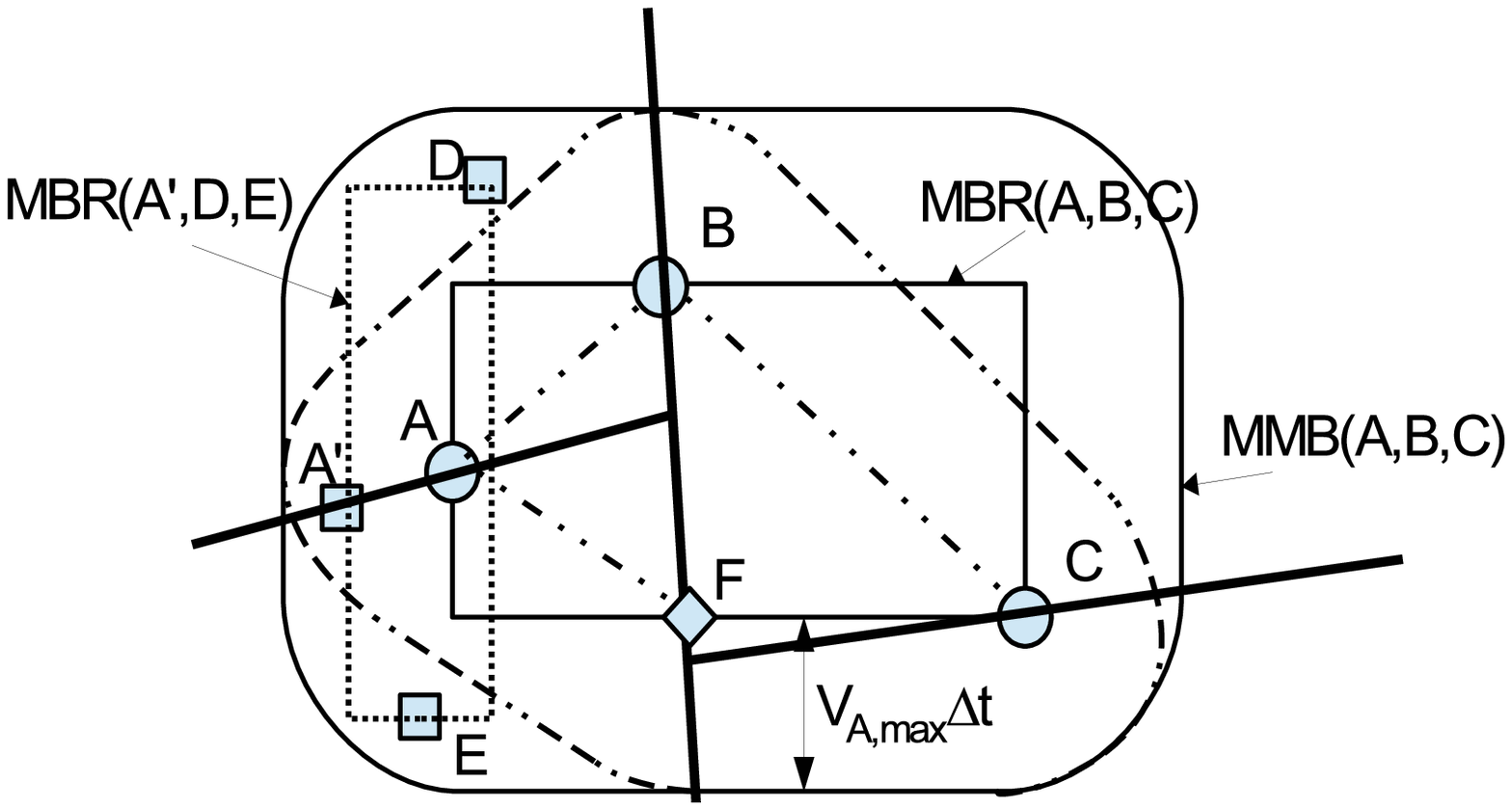, height=1.6in}
\setlength{\abovecaptionskip}{-5pt}
\setlength{\belowcaptionskip}{-10pt}
\caption{Free space vulnerability in ICliqueCloak \cite{pan2012protecting}}
\label{fig:mmb-mab-icliquecloak}
\end{figure}

A potential vulnerability of ICliqueCloak \cite{gedik2008protecting} based on Minimum Boundary Rectangle (MBR) in free space is illustrated in the Fig. \ref{fig:mmb-mab-icliquecloak}. At time $t$, user A's cloaking region is MBR(A,B,C). By Maximum Movement Boundary (MMB) constraint, at time $t+\Delta t$ , A's new location A$'$ is cloaked with say D and E. MBR(A$'$,D,E) (the dotted rectangle) is clearly covered by MMB(A,B,C), the rounded rectangle extended from MBR(A,B,C) with a radius of $v_{Amax}\times\Delta t$. The vulnerability is revealed if the attacker uses the street map (bold solid lines) and assumes the intersected points A,B,C and F as user A's possible locations at $t$. The extended convex hull of A,B,C and F by $v_{Amax}\times\Delta t$ makes A$'$ isolated from MBR(A$'$,D,E). So just as in CliqueCloak, free space assumption allows attackers to amount effective map-based attacks on ICliqueCloak.

As a consequence, employing shortest distances as spatial tolerance makes the cloaking more realistic. A$'$ and B$'$ should be called ``closed'' at $t+\Delta t$ if and only if the shortest distances from A$'$ to B$'$ and from B$'$ to A$'$ are within a certain threshold. Such a threshold is the distance each user can move in $\Delta t$. We clarify this idea in the following sections.

\subsection{Anonymization Architecture}
\begin{figure}
\centering
\epsfig{file=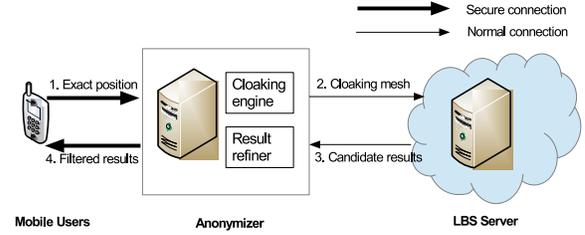, height=1.5in}
\setlength{\abovecaptionskip}{-15pt}
\setlength{\belowcaptionskip}{-10pt}
\caption{System architecture}
\label{fig:architecture}
\end{figure}

We adopt the conventional centralized trusted third party architecture which composes of mobile users, a trusted anonymizer, and an untrusted LBS service provider (see Fig. \ref{fig:architecture}). We assume the connection between the users and the anonymizer is secure and that between the anonymizer and the LBS server may be insecure. The users send periodic queries (continuous queries) to the LBS server to get location-based information. The \textit{cloaking engine} of the anonymizer in the middle effectively hides the users' exact locations using cloaking schemes. Upon receiving query results from the LBS server, the anonymizer refines and delivers only the information corresponding to the users' exact locations. Using a centralized anonymizer makes the cloaking easier as there is only one entity the users must trust, not so complicated as in peer-to-peer models. Centralized anonymizer aggregates user current locations into a global view, making the collaboration among users implicit. Also, the anonymizer mitigates the processing burden at user side.

\subsection{Spatial Tolerance via Distance Constraint}
\label{subsec:sp-tolerance}
Instead of asking users to specify the spatial tolerance in each query ($dx, dy$ for both coordinates) as in CliqueCloak, we use the maximum distance that each user can move in $\Delta t$. This kind of constraint covers a lot of real movement conditions like route direction, maximum vehicle speed, route capacity and so on. Brinkhoff's simulator \cite{brinkhoff2002framework} and other similar tools already integrate such movement conditions. 

\begin{figure}
\centering
\epsfig{file=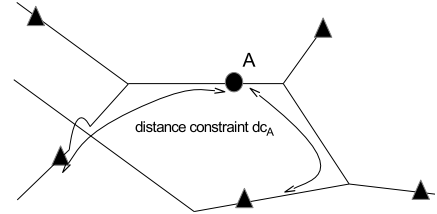, height=1.3in}
\setlength{\abovecaptionskip}{-5pt}
\caption{Reachable street segments (bounded by filled triangles) from user location (filled circle) within a distance constraint.}
\label{fig:1d-expand}
\end{figure}

From the attacker's perspective, if we know (approximately) the current location, the maximum speed of a certain user and the interval $\Delta t$ between his two consecutive queries, we can estimate the distance constraint $dc$ and compute the boundary of his movement as in Fig. \ref{fig:1d-expand}.

\subsection{Constraint Graph and Cloaking Mesh}
\label{subsec:cg-cm}
%We now introduce main notions used in MeshCloak algorithm (Section \ref{sec:algo}). 

\subsubsection{Constraint Graph}

\begin{figure}
\centering
\epsfig{file=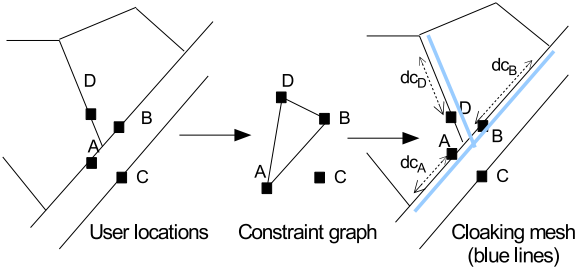, height=1.8in}
\setlength{\abovecaptionskip}{-15pt}
\setlength{\belowcaptionskip}{-10pt}
\caption{Constraint Graph and Cloaking Mesh}
\label{fig:cg-cm}
\end{figure}

To cloak users, our scheme first models user locations and their closeness relationship as an undirected and unweighted graph called \textit{constraint graph} (CG) \cite{gedik2008protecting}. The users become the nodes of CG, and an edge is created between a pair of nodes A and B if and only if the shortest distances from A to B and from B to A are smaller than both $dc_A$ and $dc_B$. In Fig. \ref{fig:cg-cm}, the locations of A,B and D are close in terms of shortest distance (i.e. A is within the boundary of D's movement and so on), so the three users are cloaked together. On the other hand, user C cannot reach A,B or D within his distance constraint and forms no edges between it and the other users. Note that CG is a \textit{dynamic} graph as user queries come and go. We show how to efficiently build the CG in Section \ref{sec:algo}.

\subsubsection{Cloaking Mesh}
After deciding a group of users to be cloaked together (called \textit{cloaking set}), the anonymizer computes the \textit{cloaking mesh}, a union of streets reachable from users within their distance constraints). For example, the cloaking set of users A,B and D has the cloaking mesh as shown on the right of Fig. \ref{fig:cg-cm}. Note that we return streets, not street segments, in cloaking meshes (see also Algorithm \ref{algo-expanding-mesh}). This change has two advantages: faster to compute and more resistant to \textit{boundary attack}. Boundary attack means the attacker tries to infer user locations using boundary points of the cloaking meshes and users' distance constraints. This kind of attack was also mentioned in ICliqueCloak's vulnerability (Fig. \ref{fig:mmb-mab-icliquecloak}).

\subsection{User Privacy}
\label{subsec:user-priv}
Following the concept of personalized location privacy \cite{gedik2008protecting}, each submitted query in MeshCloak has the following format

\begin{center}
$q(A) = \{k, t, x, y, dt, dc_A\}$
\end{center}

where $k$ is the desired minimum k-anonymity level (i.e. the cloaking set containing A should be of size at least $k$). $t$ is the timestamp of query and $(x,y)$ is the coordinates of user at $t$. Temporal tolerance is specified in $dt$, which means that the query should be processed in the interval $[t, t+dt]$, otherwise it expires and would be removed from the constraint graph. As discussed above, we set the spatial tolerance by the distance constraint $dc_A$ which is the product of query interval and user speed. For example, if user A issues a query every 10 second and its speed is at most 10m/s, then $dc_A = 10s \times 10m/s = 100m$.

\subsection{Attacker Knowledge}
We assume that the attacker wants to reidentify the location of some user A at time $t+\Delta t$ providing that he knows (approximately) A's location at time $t$ and A's maximum speed. By looking at the street map, the attacker can use A's distance constraint $dc_A$ to estimate A's movement boundary as depicted in Fig. \ref{fig:1d-expand}. Because $dc_A = v_{Amax} \times \Delta t$, the larger $\Delta t$ is, the safer A's location at time $t+\Delta t$ gets. 

To mitigate this kind of location attack, we propose to use distance constraints as spatial tolerance (Section \ref{subsec:sp-tolerance}). If a collection of users $U = \{A_i\}$, in which every user is within the distance constraints of the other users (i.e. they form a clique in the CG), then $A_i$'s location can be ``swapped'' to any other user in $U$ at time $t + \Delta t$.

As far as we know, the idea of using speed as a movement constraint first appeared in ICliqueCloak \cite{pan2012protecting}. However, ICliqueCloak considers free space. It prevents location-dependent attacks by hiding user Movement Boundary Rectangles (MBR) in two consecutive time steps. This strong assumption leads to very large cloaking areas (up to 80\% of the whole map area). Also, ICliqueCloak keeps incremental maximal cliques by processing each new edge in the CG which further increases the processing time. Table \ref{tab:algo-compare} compares our MeshCloak with CliqueCloak and ICliqueCloak in several aspects.

\begin{table}
\centering
\caption{Comparison of MeshCloak (MC) vs. CliqueCloak (CC) and ICliqueCloak (ICC)} 
\label{tab:algo-compare}
\begin{tabular}{|l|r|r|r|}
\hline
& Setting & Maximal cliques & Distance constraint  \\
\hline
MC & Map-based & Batch by Tomita & $v \times \Delta t$\\
\hline
CC \cite{gedik2008protecting} & Free-space & Local-search & Provided by users \\
\hline
ICC \cite{pan2012protecting} & Free-space & Edge-based incremental & $v \times \Delta t$\\
\hline
\end{tabular}
\setlength{\abovecaptionskip}{-5pt}
\setlength{\belowcaptionskip}{-5pt}
\end{table}

% %
\section{MeshCloak Algorithm}
\label{sec:algo}

\subsection{Precomputation of Map Distance Matrix}
\label{subsec:dist-matrix}
To be able to process a large number of queries per second, the crucial point in MeshCloak is a fast construction of the constraint graph and a fast search of maximal cliques. In our model, distance constraints determine the coverage relationship between users. As a result, given user A's location and its distance constraint $dc_A$, our goal is to quickly search for other user locations reachable from A within $dc_A$. We show below how to do this with the help of the street map.

\begin{figure}
\centering
\epsfig{file=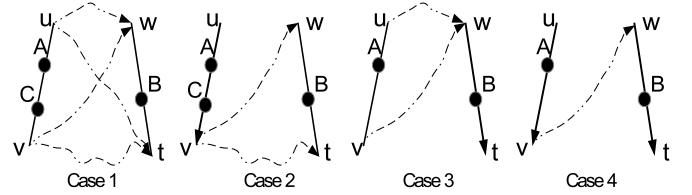, height=1.3in}
\setlength{\abovecaptionskip}{-15pt}
\setlength{\belowcaptionskip}{-10pt}
\caption{Shortest distance computation by landmarks.}
\label{fig:shortest-path}
\end{figure}

Fig. \ref{fig:shortest-path} shows possible cases of shortest distance computation by using street terminals as landmarks. To find the shortest distance $d(A,B)$ which may be different from $d(B,A)$, we utilize the fact that A and B are on certain streets. Let $d(u,v)$ be the shortest distance between two street terminals $u$ and $v$, $d(A,B)$ is computed as follows

Case 1: A and B are on two-way streets:
$d(A,B) = \min\{d(A,u) + d(u,w) + d(w,B),d(A,u) + d(u,t) + d(t,B), \\
d(A,v) + d(v,w) + d(w,B),d(A,v) + d(v,t) + d(t,B)\}$

Case 2: A (resp. B) is on a one-way (resp. two-way) street:
$d(A,B) = \min\{d(A,v) + d(v,w) + d(w,B),d(A,v) + d(v,t) + d(t,B)\}$

Case 3: A (resp. B) is on a two-way (resp. one-way) street:
$d(A,B) = \min\{d(A,u) + d(u,w) + d(w,B), d(A,v) + d(v,w) + d(w,B)\}$

Case 4: A and B are on one-way streets:
$d(A,B) = d(A,v) + d(v,w) + d(w,B)$

The other cases for two users on the same street (e.g. A and C), the computation is similar or even simpler.

From the above observations, if shortest distances between street terminals are given prior, $d(A,B)$ can be determined in $O(1)$. That is why we precompute the shortest distances between street terminals (e.g. between $u$ and $w$, $u$ and $t$ and so on). The Dijkstra algorithm is our choice. Let $V$ be the set of street terminals, $E$ be the set of streets, the shortest distances from one terminal to al the other cost $O(|E|\log |V|)$. The full computation of distance matrix $M$ will cost $O(|V||E|\log |V|)$. Note that $V$ and $E$ are map-specific information, not related to the number of users querying location services.

\begin{figure}
\centering
\epsfig{file=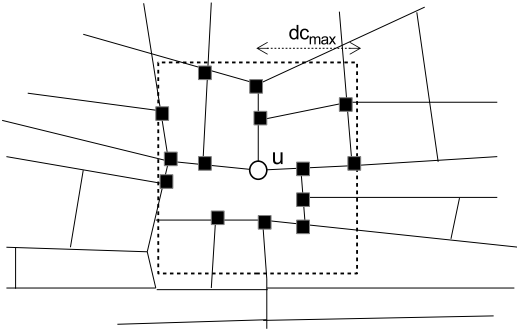, height=2.0in}
\setlength{\abovecaptionskip}{-5pt}
\setlength{\belowcaptionskip}{-10pt}
\caption{Sub-map centered at $u$ constrained by $dc_{max}$. Filled squares are terminals included in the sub-map.}
\label{fig:dijkstra}
\end{figure}

In practice, we do not need the full computation of distance matrix $M$ because of user distance constraints. Let $dc_{max}$ be the maximum distance constraint, say 1000m or 2000m, we need only a distance matrix $M(dc_{max})$ in which we retain only shortest distances upper bounded by $dc_{max}$. Fig. \ref{fig:dijkstra} illustrates this idea. The terminal $u$ and the distance $dc_{max}$ induce the sub-map $G(u,dc_{max})$ composing of terminals inside the square centered at $u$ having edge length of $2.dc_{max}$. We extract sub-maps using squares instead of circles to exploit the quadtree data structure. Now the Dijkstra algorithm runs on the sub-maps only, reducing the computation of $M(dc_{max})$ to nearly $O(|V|\log |V|)$ in which the factor $\log |V|$ is due to the quadtree search. 

All these steps are presented in Algorithm \ref{algo:dist-matrix}. We build a quadtree from the coordinates of all street terminals $V$ (Line 2). Then for each terminal $u$, a range search centered at $u$ with edge length $2.dc_{max}$ will return a set of nearby terminals $S_u$ (Lines 3-4). A subgraph and shortest distances from $u$ to other terminals in $S_u$ are implemented in Lines 5-6. Finally, we keep only tuple $(u,v,d(u,v))$ if $d(u,v) \leq dc_{max}$ (Lines 7-9).

\begin{algorithm}                      
\caption{MapDistanceMatrix}          
\label{algo:dist-matrix}                           
\begin{algorithmic}[1]                    
    \Require street map $G=(V,E)$, maximum distance constraint $dc_{max}$.
    \Ensure $M(dc_{max})$
    \State $M(dc_{max}) \leftarrow \varnothing$
    \State build the quadtree $Qt \leftarrow QuadTree(V)$
    \For{$u \in V$}
    	\State $S_u \leftarrow RangeSearch(Qt, u, dc_{max})$
    	\State $sG_u \leftarrow subgraph(G, S_u \cup \{u\}) $
    	\State $\{(u,v,d(u,v))\} \leftarrow Dijkstra(sG_u)$
    	\For{$v \in S_u$}
    		\If{$d(u,v) \leq dc_{max}$}
    			\State $M(dc_{max}) \leftarrow M(dc_{max}) \cup (u,v,d(u,v))$
    		\EndIf
    	\EndFor
    \EndFor
\end{algorithmic}
\end{algorithm}

\subsection{Building Constraint Graph}

\begin{figure}
\centering
\epsfig{file=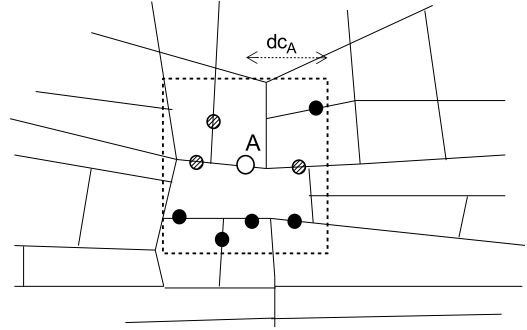, height=2.0in}
\setlength{\abovecaptionskip}{-5pt}
\setlength{\belowcaptionskip}{-10pt}
\caption{Filled circles: all potential near-by queries of $A$ constrained by $dc_A$. Dashed circles: queries truly reachable from $A$ within $dc_A$.}
\label{fig:dist-constraint}
\end{figure}

Given the distance matrix $M(dc_{max})$ and a list of queries $Q_W$ waiting for processing, we can build the constraint graph in nearly $O(|Q_W|\log |Q_W|)$ (see Algorithm \ref{algo:constraint-graph}). Again, we apply the idea of filtering out far-away queries by using the quadtree structure. In Fig. \ref{fig:dist-constraint}, a range search centered at A with edge length $2.dc_A$ may return eight potential queries (denoted as filled or dashed circles). Combining with $M(dc_{max})$, we further eliminate queries not truly reachable from A within $dc_A$ (the filled circles), retaining only three queries (denoted as dashed circles). Line 6 runs in $O(1)$ thanks to case-by-case checking described in Section \ref{subsec:dist-matrix}. Note that, we keep only undirected edges in $CG(Q_W)$ (Line 9).

\begin{algorithm}                      
\caption{ConstraintGraph}          
\label{algo:constraint-graph}                           
\begin{algorithmic}[1]                    
    \Require street map $G=(V,E)$, distance matrix $M(dc_{max})$, list of queries $Q_W$
    \Ensure constraint graph $CG(Q_W)$
    \State $CG(Q_W) \leftarrow \varnothing$
    \State build the quadtree $Qt \leftarrow QuadTree(Q_W)$
    \For{$q \in Q_W$}
    	\State $S_q \leftarrow RangeSearch(Qt, q, dc_q)$
    	\For{$q' \in S_q$}
    		\State $d(q,q') \leftarrow  shortest\;distance(q, q', M(dc_{max}))$
    		\If{$d(q,q') \leq dc_q$}
    			\State $CG(Q_W) \leftarrow CG(Q_W) \cup (q, q')$
    		\EndIf
    	\EndFor
    \EndFor
    \State keep only edge $(q,q')$ iff $(q',q)$ exists in $CG(Q_W)$
\end{algorithmic}
\end{algorithm}

\subsection{Cloaking Mesh}
Given a clique $U$ of users in the CG, the cloaking mesh (Fig. \ref{fig:cg-cm}) of $U$ can be computed as the union of streets reachable from any user A in $U$ within $dc_A$ (see Fig. \ref{fig:1d-expand}). 

\begin{center}
$CMesh(U) = \bigcup_{A \in U} EMesh(A)$
\end{center}

For each successfully cloaked user A, we compute its \textit{Expanding Mesh} EMesh(A) as in Algorithm \ref{algo-expanding-mesh}. First, we identify the map street $e$ that A belongs to (Line 1). We initialize the empty EMesh(A), an array \textit{visited} and a queue $q$ in Lines 2-3. Depending on $e$'s direction, one or two items are enqueued into $q$ (Lines 4-8). Each item is a tuple of $(u,v,L)$ which means we examine the directed edge from $u$ to $v$ and the remaining length is $L$. Finally, EMesh(A) is updated incrementally by breadth-first-search on $G(V,E)$ (Lines 9-15) as long as the remaining length $L$ is not negative.

\begin{algorithm}
\caption{ExpandingMesh}
\label{algo-expanding-mesh}
\begin{algorithmic}[1]
	\Require street map $G=(V,E)$, user A's location $(x,y)$, distance constraint $dc_A$
	\Ensure expanding mesh $EMesh(A)$
	\State $e(u,v)$: the street that $(x,y)$ belongs to
	\State $EMesh(A) \leftarrow \{e\}$, $visited(u) = False \;\;\forall u \in  V$
	\State queue $q \leftarrow \varnothing$
	\If{$e$ is one-way}
		\State $q.enqueue((u,v,L=dc_A))$, $visited(v) = True$
	\Else{  // two-way street}
		\State $q.enqueue((u,v,L=dc_A))$, $visited(v) = True$
		\State $q.enqueue((v,u,L=dc_A))$, $visited(u) = True$, 
	\EndIf
	\While {$q$ not empty}
		\State $e(u,v,L) \leftarrow q.dequeue()$
		\If {$e.L > 0$}
			\For{$w \in N(v) \text{ AND } visited(w) == False $}
				\State $EMesh(A) \leftarrow EMesh(A) \cup (v,w)$
				\State $q.enqueue((v,w,L=e.L - d(v,w)))$
				\State $visited(w) = True$
			\EndFor
		\EndIf
	\EndWhile		
	\State \Return $EMesh(A)$
\end{algorithmic}
\end{algorithm}

\subsection{Cloaking Algorithm}
As in CliqueCloak, incoming queries are stored in a queue. Each query has one of four possible states: state NEW if the query is a newly arrived, state EXPIRED if the query cannot be cloaked in its interval $[t, t+dt]$, state WAITING if query is waiting for cloaking and does not expire, state SUCCEEDED if query is successfully cloaked with some other queries. We denote $Q_N$,$Q_E$,$Q_W$ and $Q_S$ as the sets of NEW, EXPIRED, WAITING and SUCCEEDED queries respectively.

Unlike per-query sequential processing in CliqueCloak, our MeshCloak processes incoming queries in small batches, one batch per second (see Algorithm \ref{algo:meshcloak}). In each batch, MeshCloak involves four steps: removing expired queries (Lines 3-7), collecting new queries and building the constraint graph (Lines 8-10), listing all maximal cliques (Line 11) and processing successfully cloaked queries (Lines 12-17). Temporal tolerance checking is carried out in Line 5. K-anonymity level of the query $q$ is checked in Line 14.

Note that each batch of queries must be processed fast enough to prevent as much as possible query expiration. That is the reason for using Tomita algorithm \cite{tomita2006worst}, one of the fastest all-maximal-cliques listing methods. Processing queries one-by-one as in CliqueCloak \cite{gedik2008protecting} and ICliqueCloak \cite{pan2012protecting} incurs much higher time for maintaining maximal cliques. 

\begin{algorithm}                      
\caption{MeshCloak}          
\label{algo:meshcloak}                           
\begin{algorithmic}[1]                    
    \Require street map $G=(V,E)$, maximum distance constraint $dc_{max}$, a stream of queries $Q$.
    \Ensure cloaking results for queries
    \State $M(dc_{max}) \leftarrow MapDistanceMatrix(G)$.
    \While{cloaking engine is running}
    	\State // remove expired queries from $Q_W$, add them to $Q_E$
    	\For{query $q \in Q_W$}
    		\If{$q.t + q.dt < now$}
    			\State $Q_E \leftarrow Q_E \cup \{q\}$
	    		\State $Q_W \leftarrow Q_W - \{q\}$
	    	\EndIf
    	\EndFor
    	\State collect queries in the last second from $Q$ into $Q_N$
    	\State $Q_W \leftarrow Q_W \cup Q_N$
    	\State $CG(Q_W) \leftarrow ConstraintGraph(Q_W, M(dc_{max}))$
    	\State $CL \leftarrow $ all maximal cliques in $CG(Q_W)$ by \cite{tomita2006worst}
    	\For{query $q \in Q_W$}
    		\State $CL(q) \leftarrow \arg\max_{c} \{|c|\;\;| c\in CL , c \ni q\}$
    		\If{$|CL(q)| \ge q.k$}
    			\State output $CloakingMesh(CL(q))$
    			\State $Q_S \leftarrow Q_S \cup \{q\}$
    			\State $Q_W \leftarrow Q_W - \{q\}$
    		\EndIf
    	\EndFor
    \EndWhile
\end{algorithmic}
\end{algorithm}

% %
\section{Evaluation}
\label{sec:evaluation}

In this section, we evaluate the efficiency and effectiveness of our scheme MeshCloak. The efficiency is measured in cloaking time per request while the effectiveness is evaluated in terms of success rate, average mesh length, relative k-anonymity and relative temporal tolerance. Experimental setting is first described in Section \ref{subsec:exp-setting}, followed by experimental results in \ref{subsec:exp-result}. All cloaking algorithms are implemented in C++ and run on a desktop PC with $Intel^{\circledR}$ Core i7-6700@ 3.4Ghz, 16GB memory.

\subsection{Experimental Setting}
\label{subsec:exp-setting}
Real road maps from many cities \footnote{https://www.openstreetmap.org/} can be extracted before being input to Brinkhoff's simulator \cite{brinkhoff2002framework}. We test five real maps: Oldenburg (Germany), Hanoi (Vietnam), Paris-zone1 (France), Singapore and San Joaquin (USA). The characteristics of five datasets are summarized in Table \ref{tab:map-prop}. 

\begin{table*}[!t]
\centering
\caption{Map properties} \label{tab:map-prop}
\begin{tabular}{|l|r|r|r|r|r|}
\hline
& \textbf{Oldenburg} &\textbf{Hanoi} & \textbf{Paris-zone1} & \textbf{Singapore} & \textbf{San Joaquin} \\
\hline
Nodes & 6,105 & 27,213 & 42,494 & 54,674 & 52,528 \\
\hline
Edges & 7,029 & 31,562 & 63,722 & 74,053 & 57,284 \\
\hline
Width(km) & 23.57 & 11.97 & 18.87 & 33.65 & 22.97 \\
\hline
Height(km) & 26.92 & 12.48 & 10.20 & 16.14 & 20.19 \\
\hline
Area($km^2$) & 634.44 & 149.40 & 192.47 & 543.11 & 463.76 \\
\hline
Total street length(km) & 1,301.70 & 1,531.64 & 4,344.18 & 5,856.25 & 2,853.83 \\
\hline
%Avg.node-node-distance(km) & 0.185 & 1.705\\ 
%\hline
\end{tabular}
\setlength{\abovecaptionskip}{-5pt}
\setlength{\belowcaptionskip}{-5pt}
\end{table*}

We customized Brinkhoff's simulator to incorporate real user speed and query interval (Table \ref{tab:default-exp}). We simulate 100,000 users moving according to two speed profiles P1 and P2. Each user issues 11 queries with personalized k-anonymity. We test two settings of k-anonymity: $k \in [2-5]$ and $k \in [2-10]$. The first query time of each user is assigned with a random integral value in the range [0 - 50].

In Table \ref{tab:default-exp}, \textit{speed proportion} indicates the proportions of users moving with given speeds. For example, in P1 there are 25\% of users moving with speed 10m/s. Similarly, \textit{query interval} and \textit{query interval proportion} let us know the proportions of users issuing queries after a given time interval. For example, in P1 there are 50\% of users issuing one query every 5 seconds. As stated in Section \ref{subsec:user-priv}, user speed and query interval define the distance constraint. Finally, temporal tolerance should not exceed the minimum query interval. For each profile, we test different temporal tolerance values.

The actual clique size that a user belongs to may be larger than the user's k-anonymity. Similarly, users may be cloaked sooner than the time constrained by the temporal tolerance. As a result, we define \textit{relative k-anonymity} (\textit{rel.k}) as the ratio of the actual clique size to user's desired k-anonymity and \textit{relative temporal tolerance} (\textit{rel.dt}) as the ratio of the actual cloaking delay to user's temporal tolerance $dt$.

\begin{table}
\centering
\caption{Experiment settings} \label{tab:default-exp}
\begin{tabular}{|l|p{5cm}|}
\hline
\textbf{Parameter} &\textbf{Values} \\
\hline
No. users & 100,000 \\
\hline
No. queries/user & 11 \\
\hline
First query time & [0-50s] \\
\hline
k-anonymity & [2-5], [2-10] \\
\hline
\textbf{Speed profile 1 (P1)} &  \\
\hline
Speed (m/s) & [10, 20, 30, 50] \\
\hline
Speed proportion & [0.25, 0.25, 0.25, 0.25] \\
\hline
Query interval  & [5s, 10s, 20s] \\
\hline
Query interval proportion & [0.5, 0.3, 0.2] \\
\hline
Temporal tolerance & [3s, 4s, 5s] \\
\hline
\textbf{Speed profile 2 (P2)} &  \\
\hline
Speed (m/s) & [10, 20, 30, 50] \\
\hline
Speed proportion & [0.25, 0.25, 0.25, 0.25] \\
\hline
Query interval  & [20s, 30s] \\
\hline
Query interval proportion & [0.5, 0.5] \\
\hline
Temporal tolerance & [3s, 5s, 7s, 10s] \\
\hline
\end{tabular}
\setlength{\abovecaptionskip}{-5pt}
\setlength{\belowcaptionskip}{-5pt}
\end{table}

\subsection{Experimental Results}
\label{subsec:exp-result}

\subsubsection{Query Volume}
\begin{figure*}[t!]
 	\centering
         \begin{subfigure}[b]{0.47\textwidth}
                 \centering
                 \epsfig{file=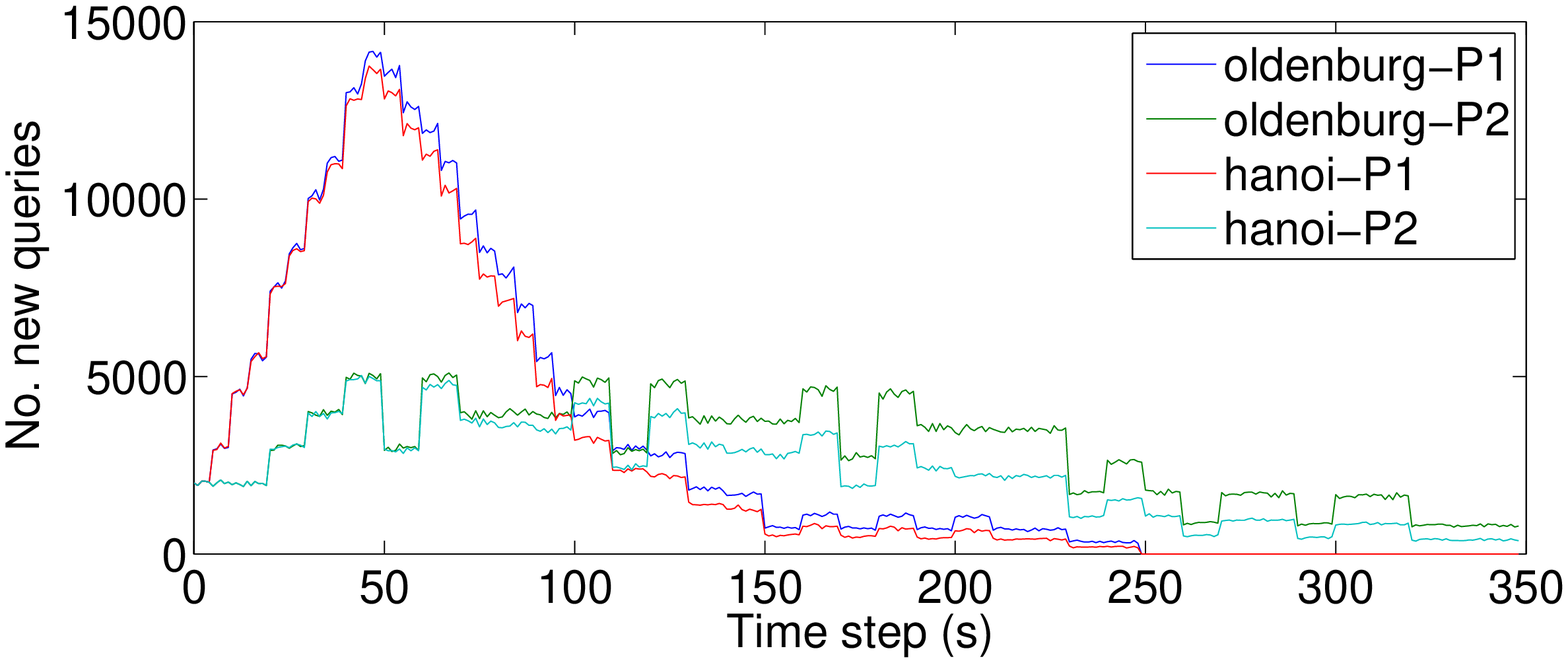, height=1.5in}
                 \caption{Number of NEW queries: Oldenburg - Hanoi}	
                 \label{fig:vol-new}
         \end{subfigure}
         \hfill
         \begin{subfigure}[b]{0.47\textwidth}
                 \centering
                 \epsfig{file=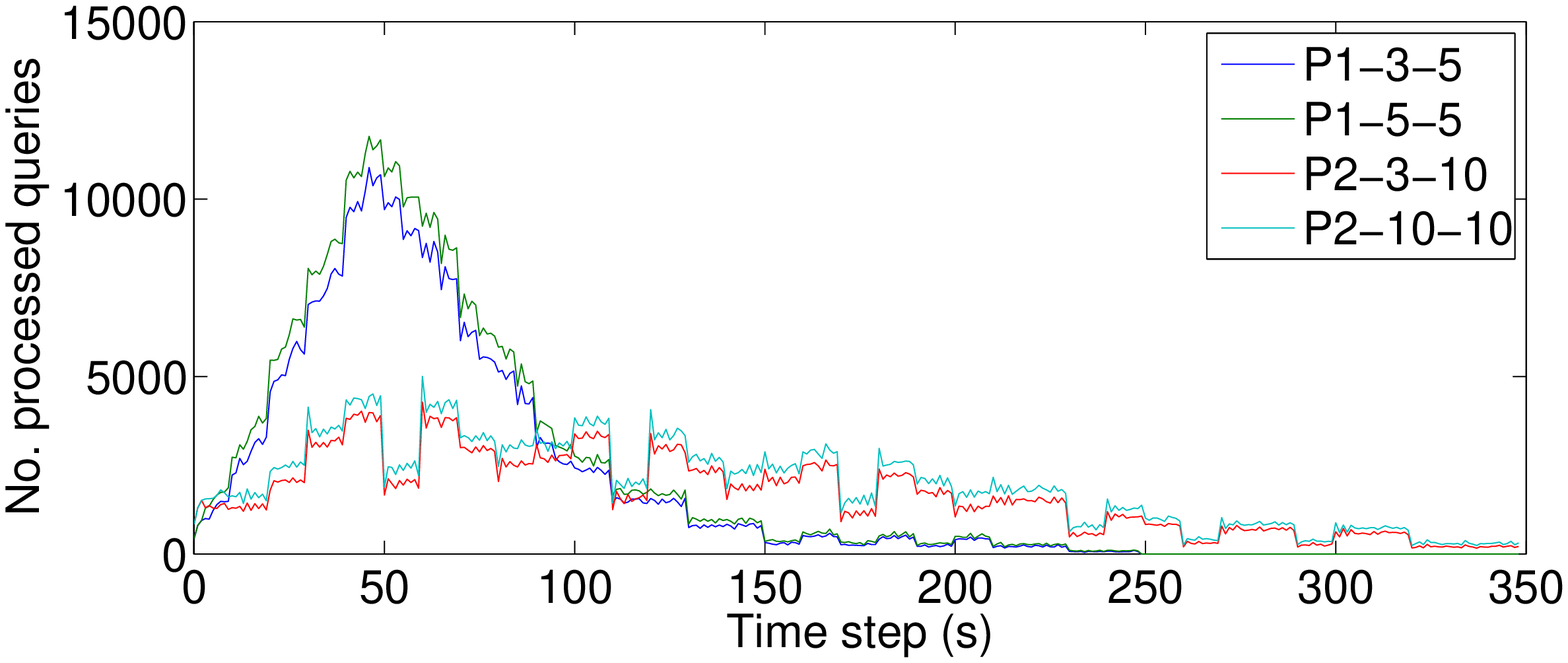, height=1.5in}
                 \caption{Number of SUCCEEDED queries: Paris-zone1}
                 \label{fig:vol-succ}
         \end{subfigure}
         \hfill         
         \begin{subfigure}[b]{0.47\textwidth}
                 \centering
                 \epsfig{file=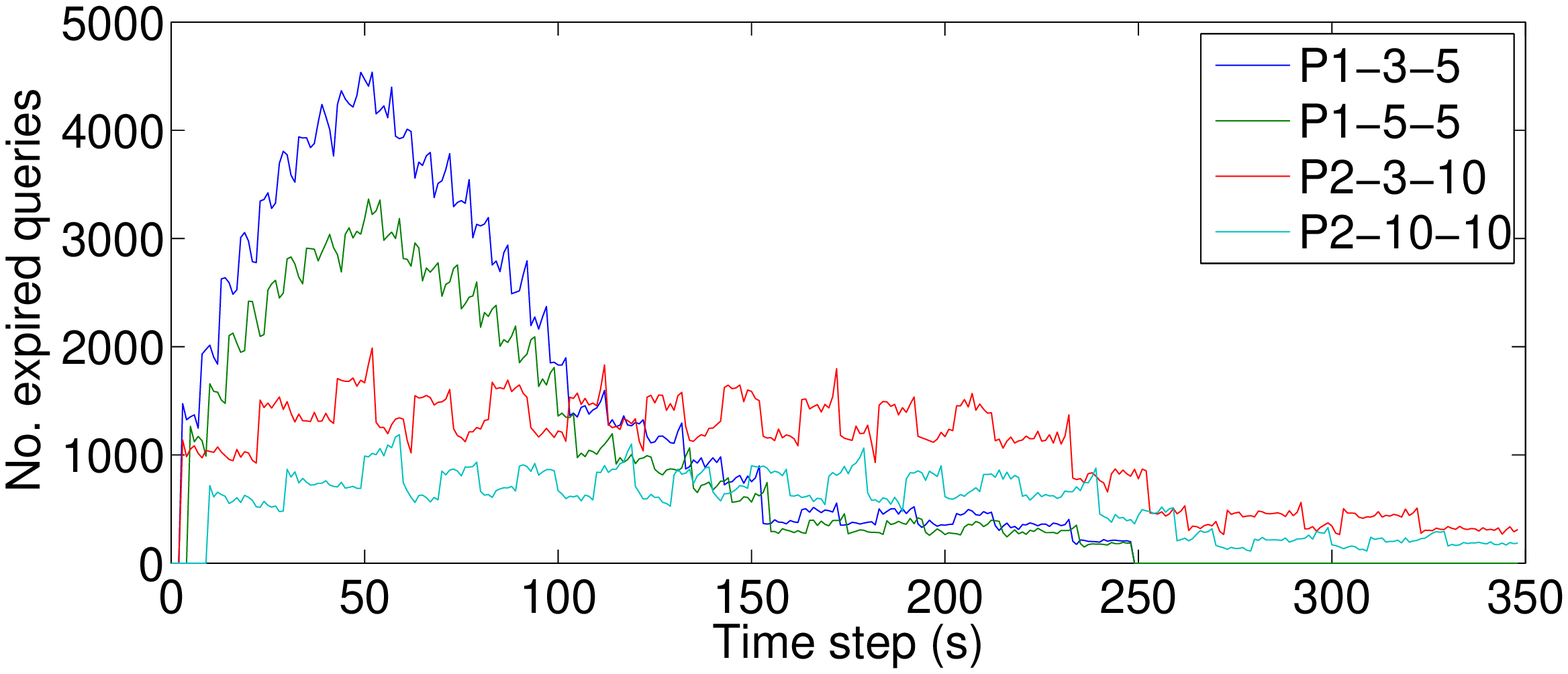, height=1.5in}
                 \caption{Number of EXPIRED queries: Singapore}
                 \label{fig:vol-exp}
         \end{subfigure}
         \hfill         
         \begin{subfigure}[b]{0.47\textwidth}
                  \centering
                  \epsfig{file=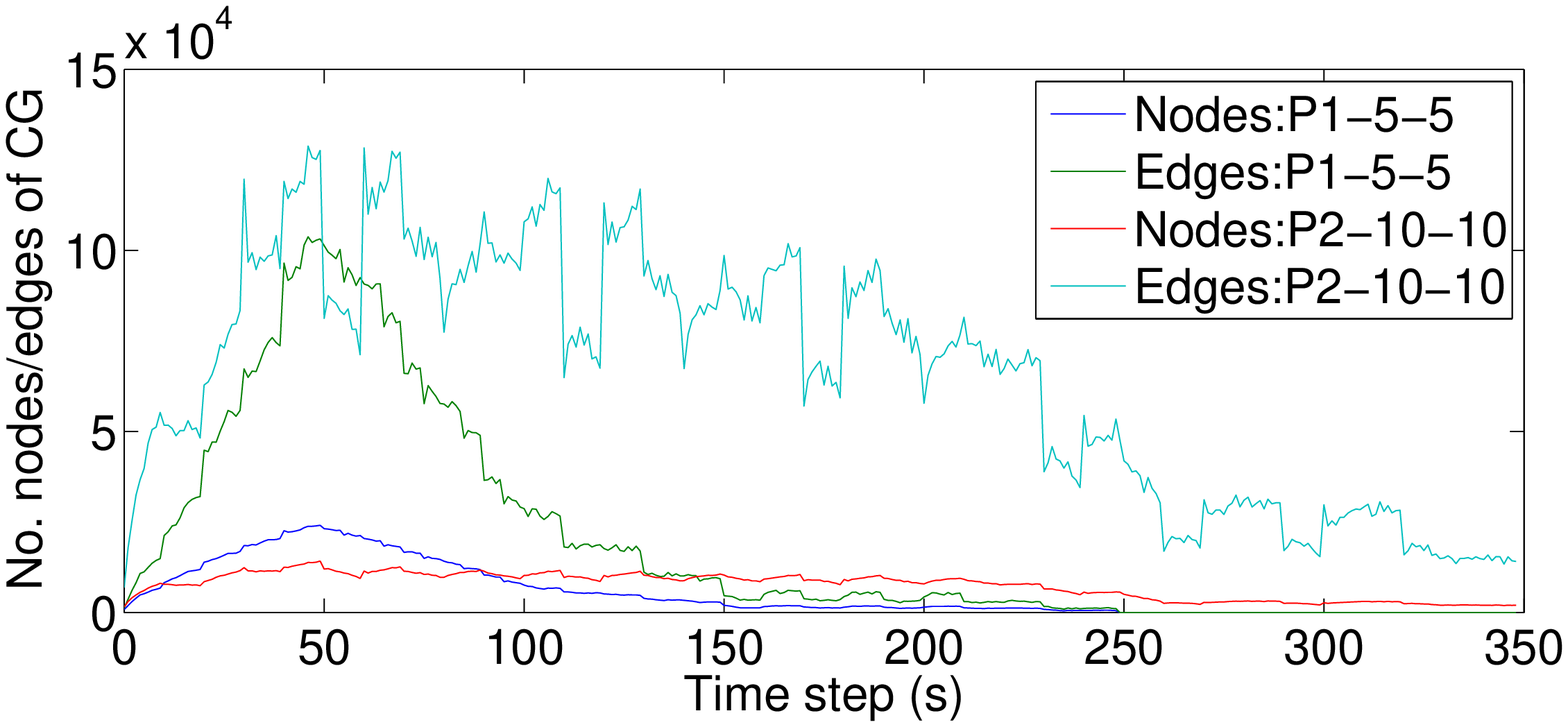, height=1.5in}
                 \caption{Number of nodes/edges of CG: San Joaquin}
                  \label{fig:vol-node}
         \end{subfigure}        	

     \caption{Query volume (P1-3-5 denotes the configuration of P1 speed profile, $dt=3$, $k=5$) and so on}
     \label{fig:query-volume}
\end{figure*}

We visualize the query volume at different time steps in Fig. \ref{fig:query-volume}. The query volume comprises the numbers of NEW, SUCCEEDED, EXPIRED and WAITING queries. Fig. \ref{fig:vol-new} shows the volume of NEW queries on Oldenburg and Hanoi maps with profiles P1 and P2. Clearly, P2 provides flatter volume progression. This is explained by P2's query interval which includes only two values 20s and 30s. The similar tendencies are observed in the volume of SUCCEEDED and EXPIRED queries. These curves imply fairly stable success rates which are defined as the ratio of the number of SUCCEEDED queries to the total number of SUCCEEDED and WAITING queries. In Fig. \ref{fig:vol-node}, we show how the number of nodes and edges of the Constraint Graph change as time passes. The number of nodes (edges) in the CG may rise up to 25,000 (130,000) but the Tomita algorithm still needs a fraction of time to list all the maximal cliques.

\subsubsection{Effectiveness and efficiency}
\begin{figure*}[t!]
 	\centering
         \begin{subfigure}[b]{0.23\textwidth}
                 \centering
                 \epsfig{file=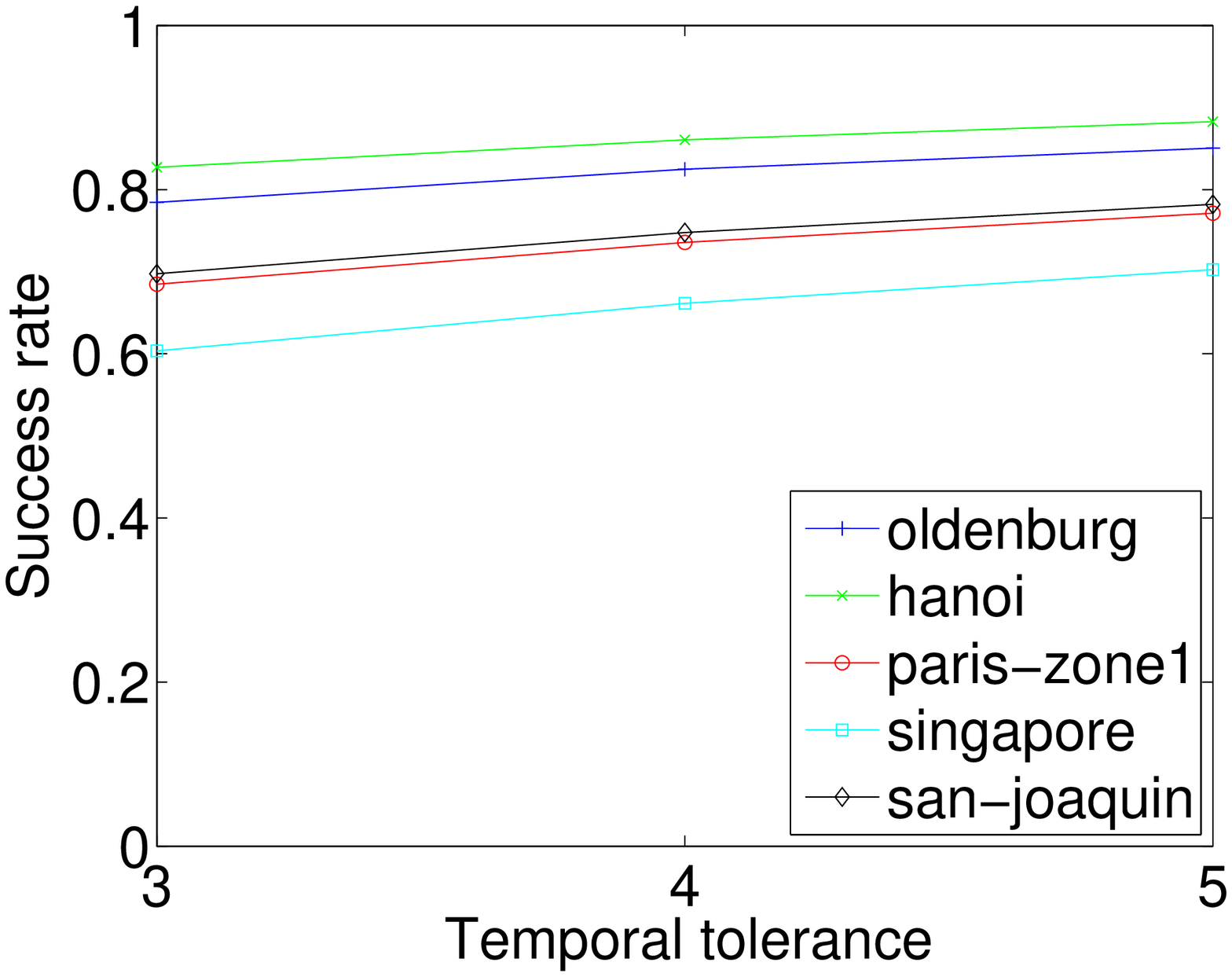, height=1.35in}
                 \setlength{\abovecaptionskip}{-10pt}
                 \caption{P1, k=[2-5]}	
%                 \label{fig:util-PL-1}
         \end{subfigure}
         \hfill
         \begin{subfigure}[b]{0.23\textwidth}
                 \centering
                 \epsfig{file=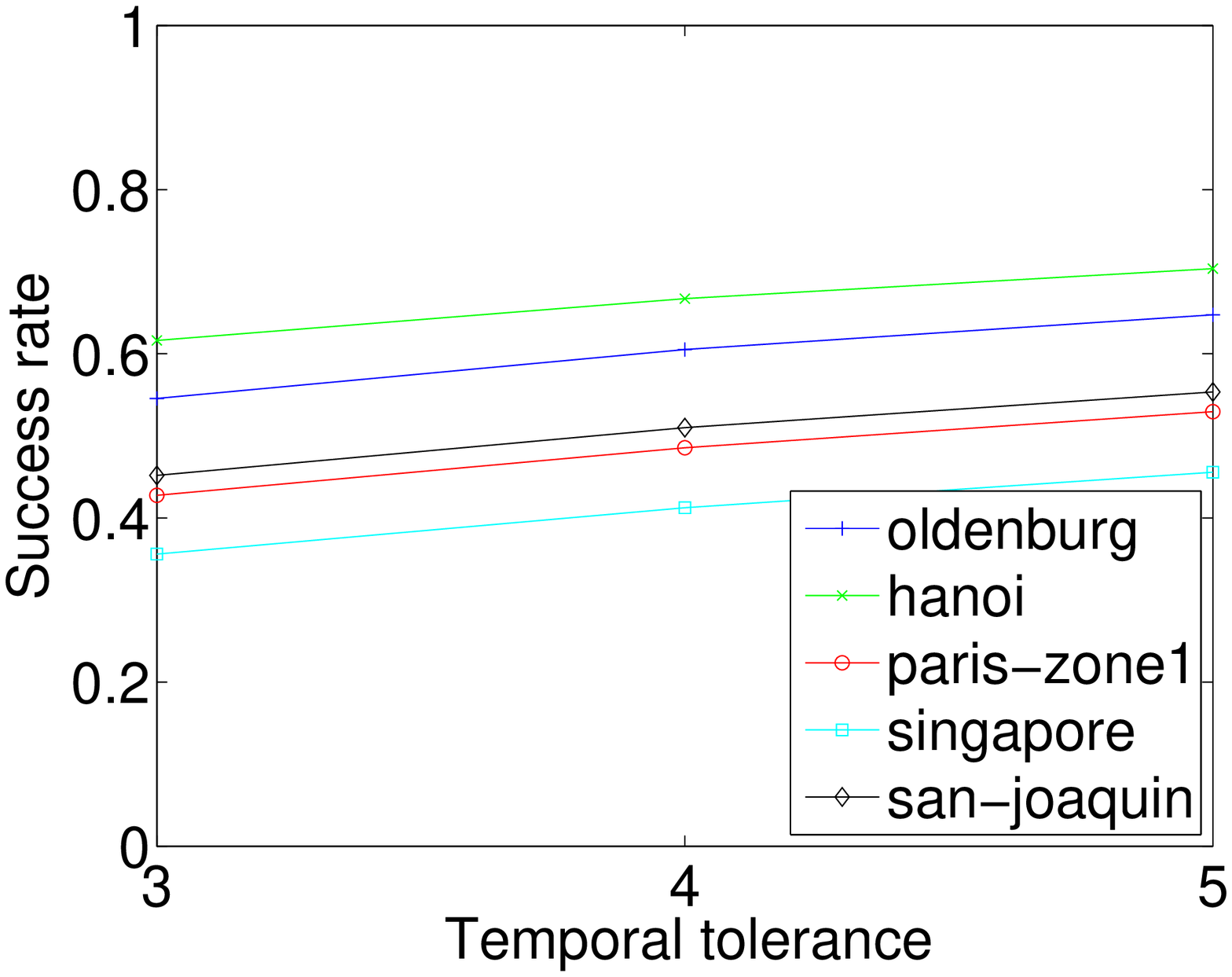, height=1.35in}
                 \setlength{\abovecaptionskip}{-10pt}
                 \caption{P1, k=[2-10]}
%                 \label{fig:util-CC-1}
         \end{subfigure}
         \hfill         
         \begin{subfigure}[b]{0.23\textwidth}
                 \centering
                 \epsfig{file=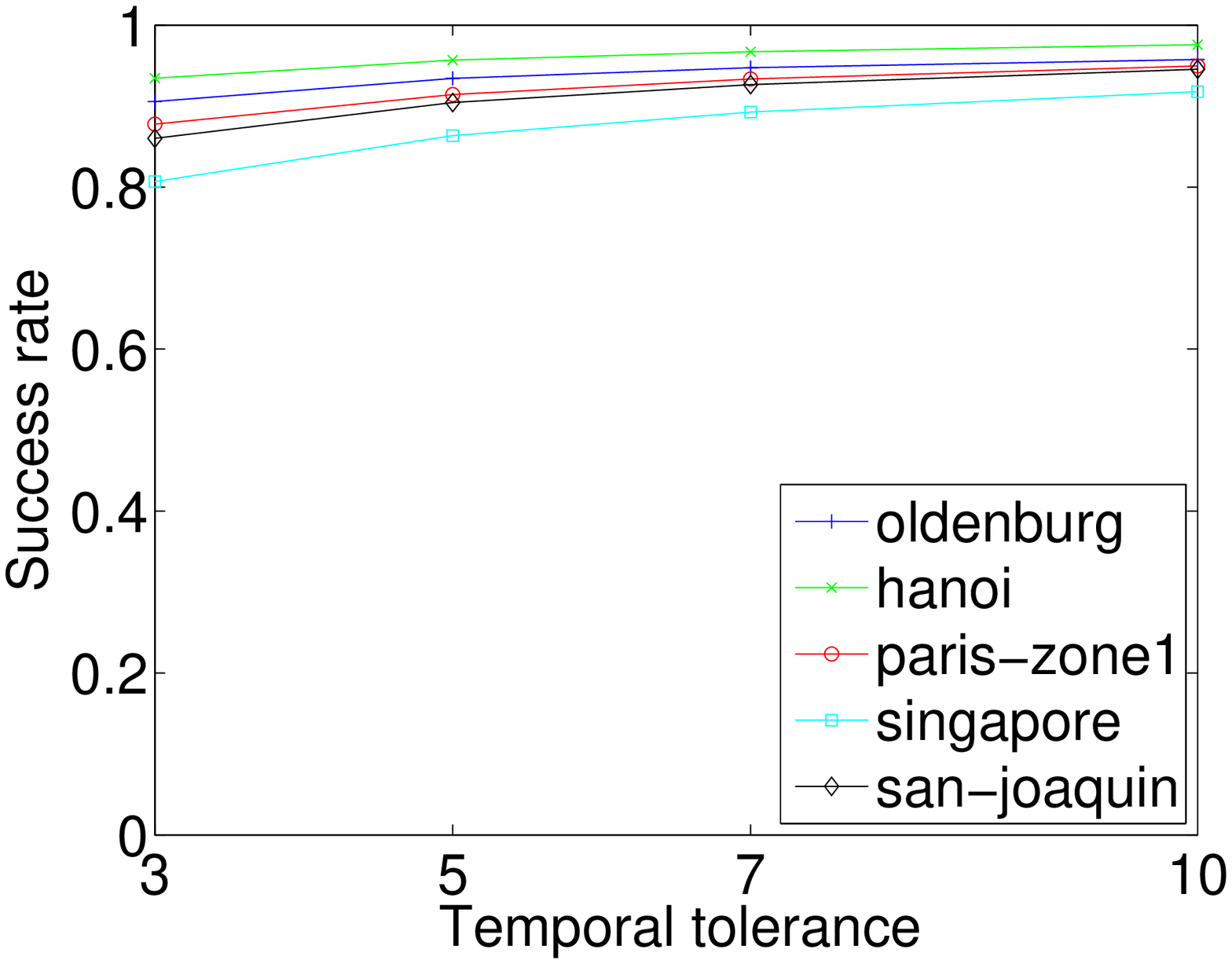, height=1.35in}
                 \setlength{\abovecaptionskip}{-10pt}
                 \caption{P2, k=[2-5]}
%                 \label{fig:util-APD-1}
         \end{subfigure}
         \hfill         
         \begin{subfigure}[b]{0.23\textwidth}
                  \centering
                  \epsfig{file=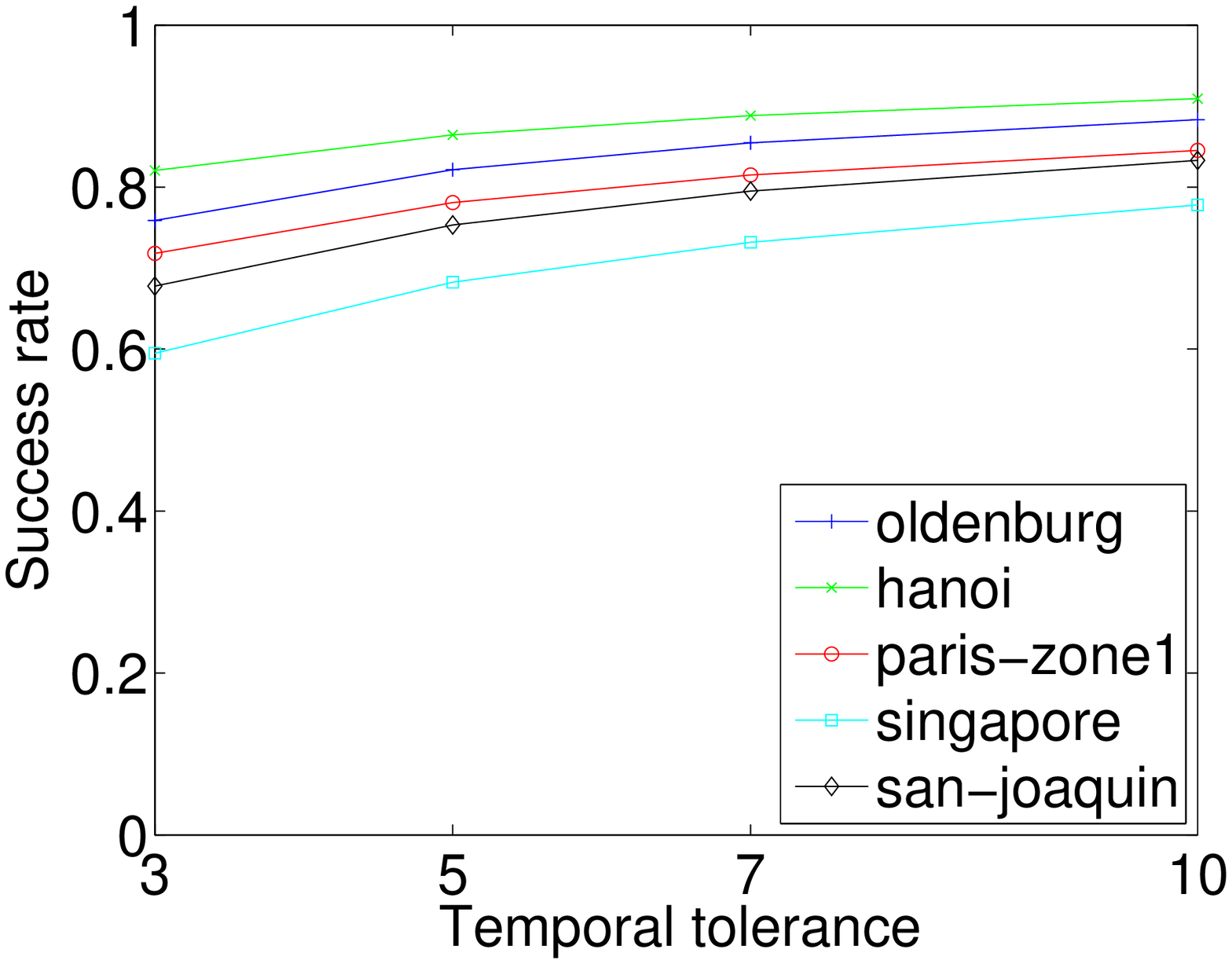, height=1.35in}
                  \setlength{\abovecaptionskip}{-10pt}
                 \caption{P2, k=[2-10]}
%                  \label{fig:util-Dist-1}
         \end{subfigure}        	

     \caption{Success rate}
     \label{fig:succ-rate}
\end{figure*}

\begin{figure*}[t!]
 	\centering
         \begin{subfigure}[b]{0.23\textwidth}
                 \centering
                 \epsfig{file=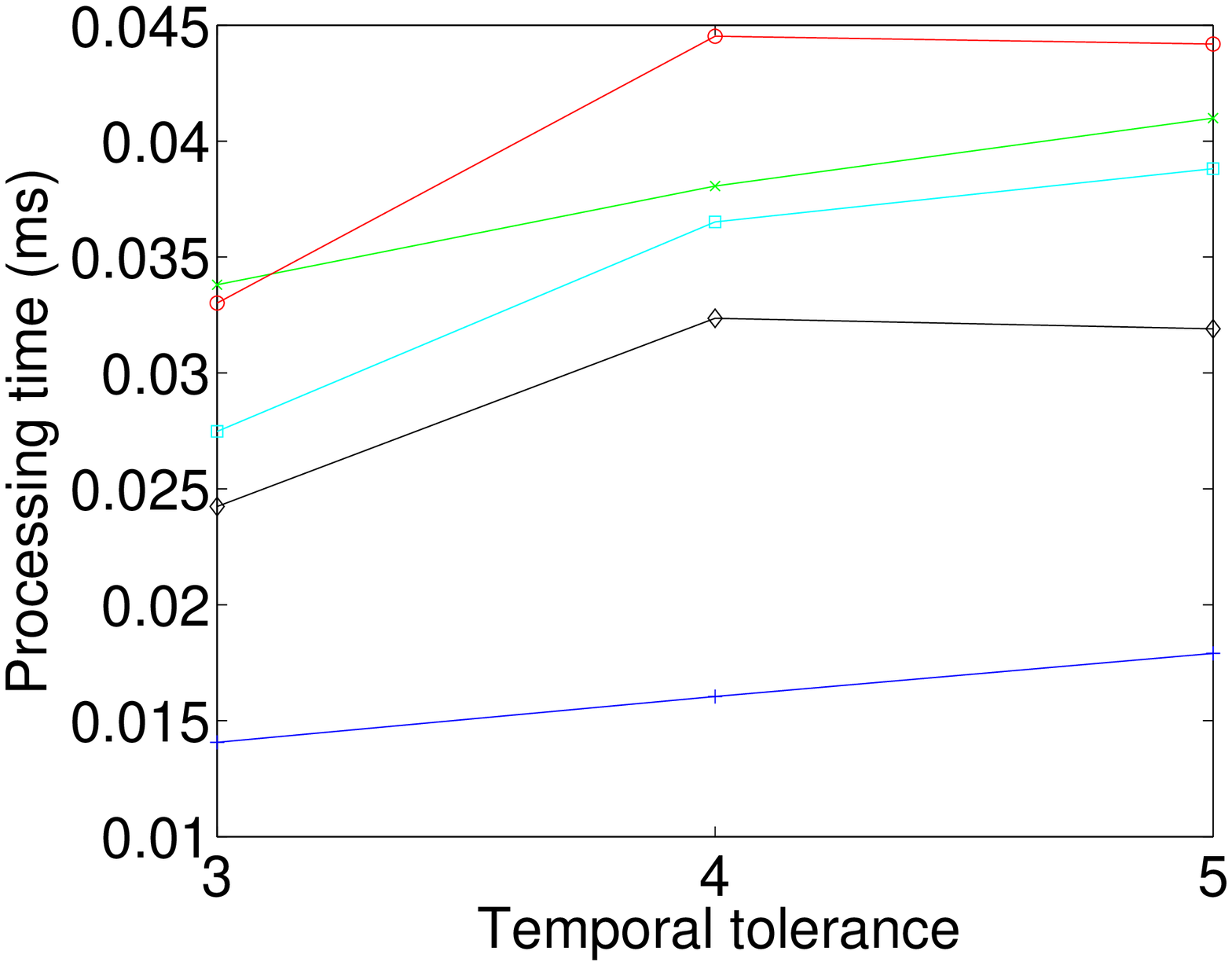, height=1.35in}
                 \setlength{\abovecaptionskip}{-10pt}
                 \caption{P1, k=[2-5]}	
%                 \label{fig:util-PL-1}
         \end{subfigure}
         \hfill
         \begin{subfigure}[b]{0.23\textwidth}
                 \centering
                 \epsfig{file=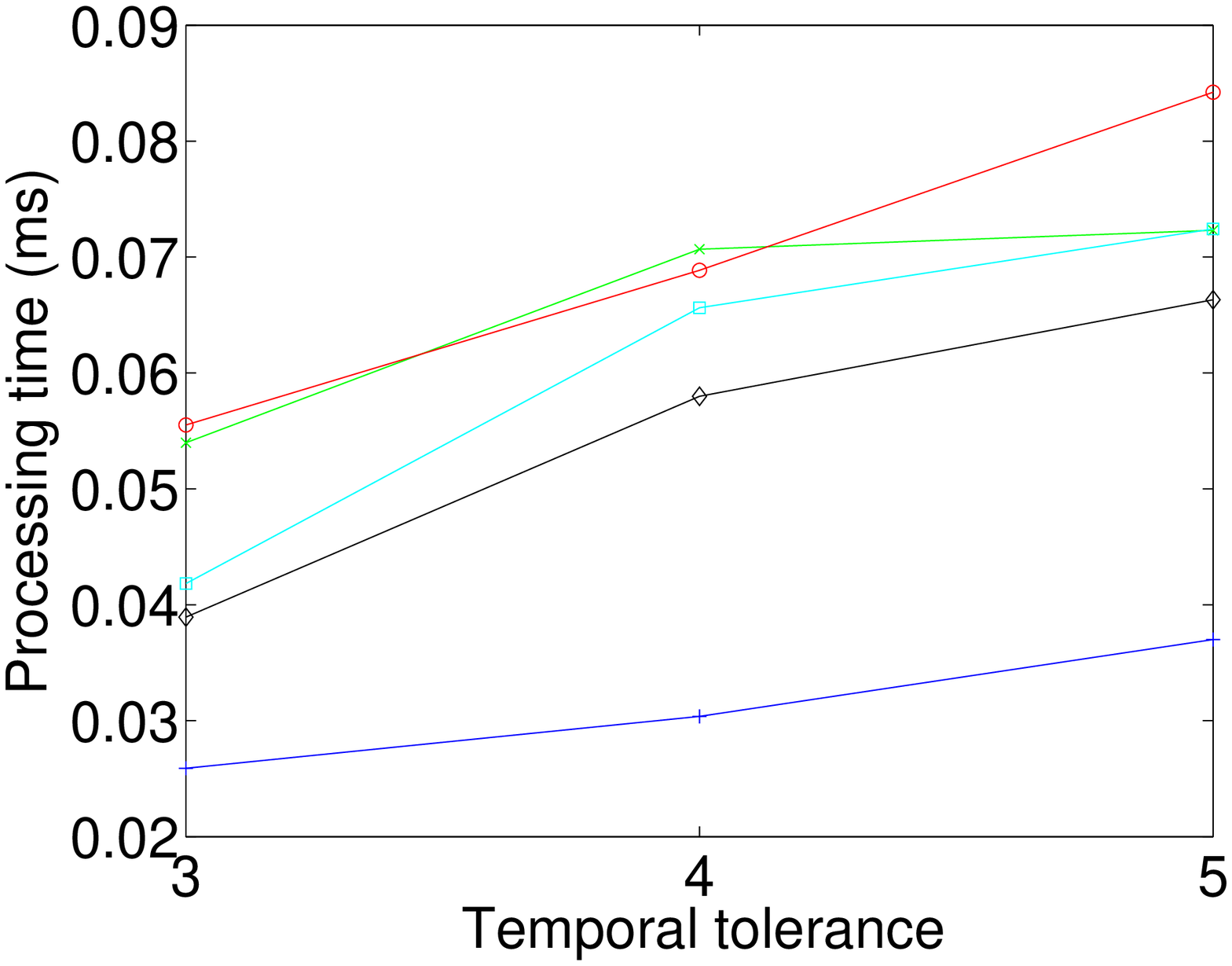, height=1.35in}
                 \setlength{\abovecaptionskip}{-10pt}
                 \caption{P1, k=[2-10]}
%                 \label{fig:util-CC-1}
         \end{subfigure}
         \hfill         
         \begin{subfigure}[b]{0.23\textwidth}
                 \centering
                 \epsfig{file=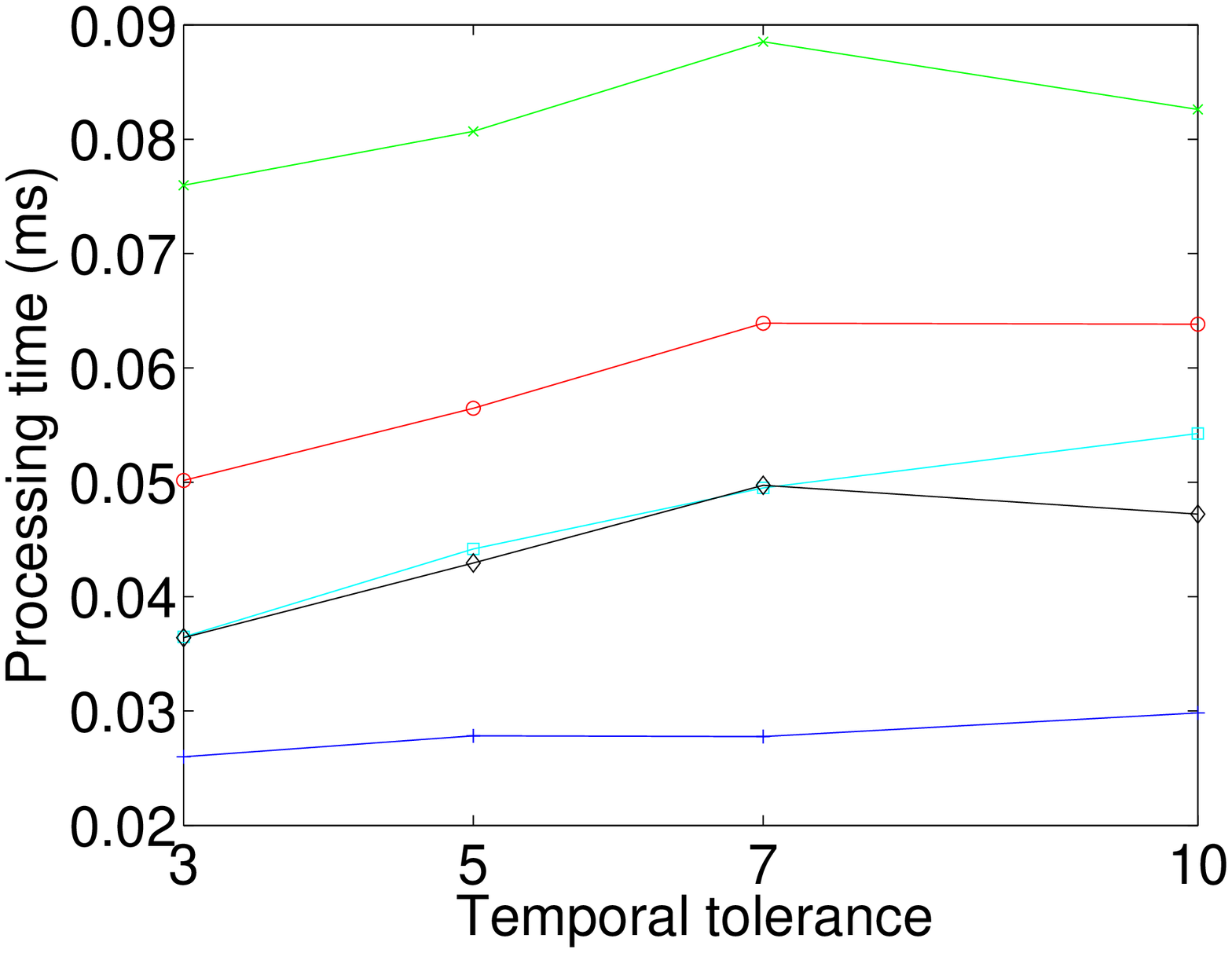, height=1.35in}
                 \setlength{\abovecaptionskip}{-10pt}
                 \caption{P2, k=[2-5]}
%                 \label{fig:util-APD-1}
         \end{subfigure}
         \hfill         
         \begin{subfigure}[b]{0.23\textwidth}
                  \centering
                  \epsfig{file=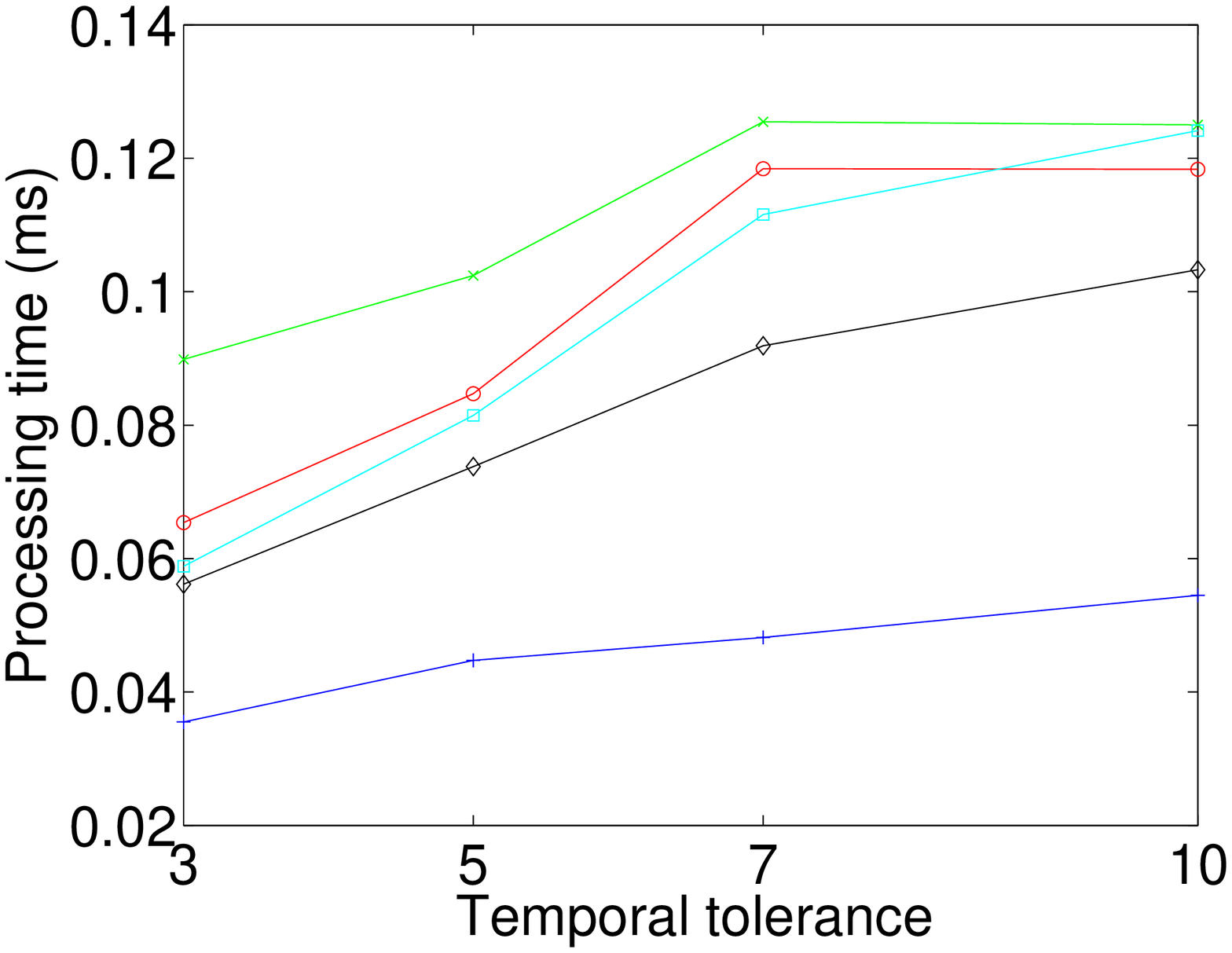, height=1.35in}
                  \setlength{\abovecaptionskip}{-10pt}
                 \caption{P2, k=[2-10]}
%                  \label{fig:util-Dist-1}
         \end{subfigure}        	

     \caption{Processing time per query (the same legends as in Fig. \ref{fig:succ-rate})}
     \label{fig:proc-time}
\end{figure*}

\begin{figure*}[t!]
 	\centering
         \begin{subfigure}[b]{0.23\textwidth}
                 \centering
                 \epsfig{file=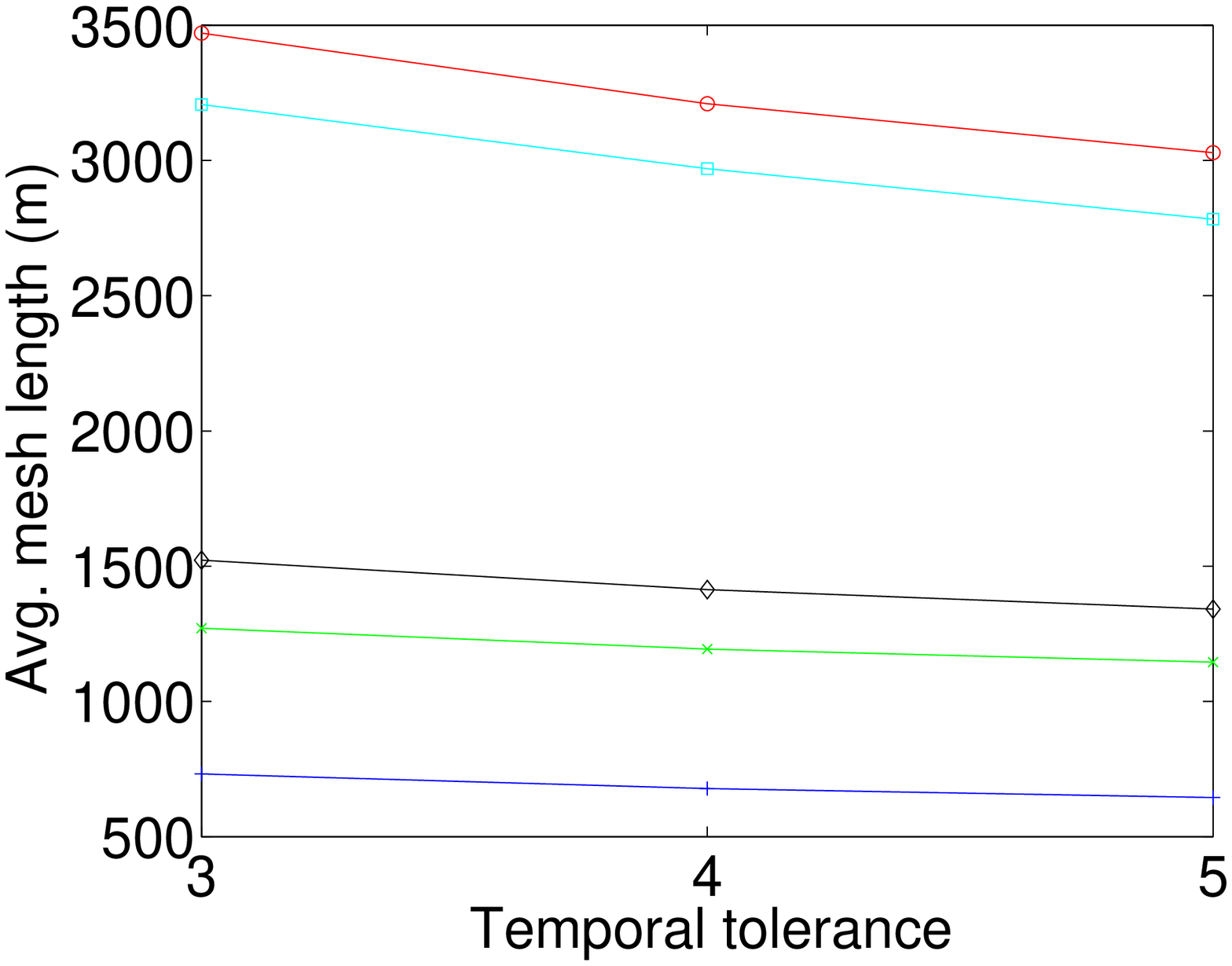, height=1.35in}
                 \setlength{\abovecaptionskip}{-10pt}
                 \caption{P1, k=[2-5]}	
%                 \label{fig:util-PL-1}
         \end{subfigure}
         \hfill
         \begin{subfigure}[b]{0.23\textwidth}
                 \centering
                 \epsfig{file=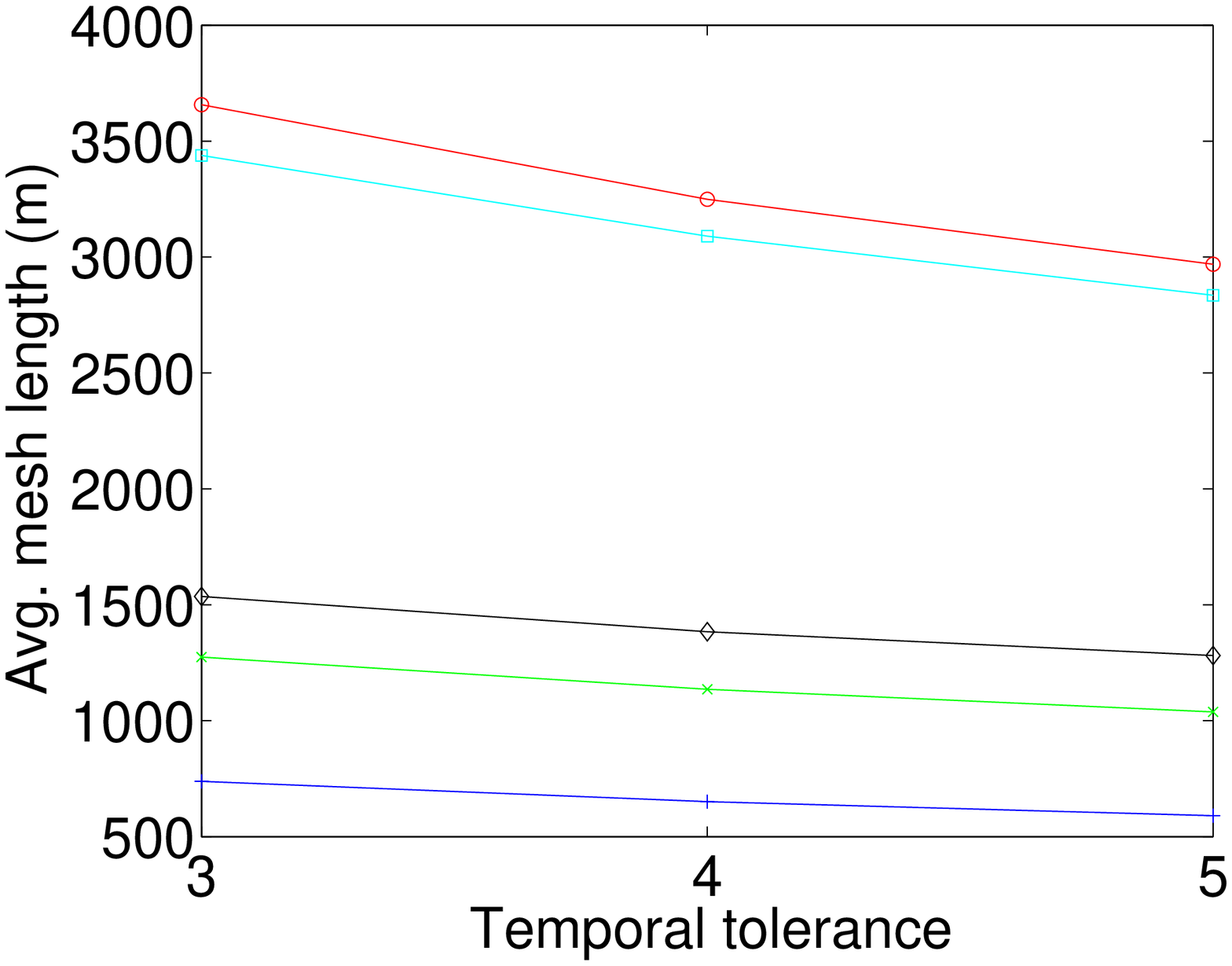, height=1.35in}
                 \setlength{\abovecaptionskip}{-10pt}
                 \caption{P1, k=[2-10]}
%                 \label{fig:util-CC-1}
         \end{subfigure}
         \hfill         
         \begin{subfigure}[b]{0.23\textwidth}
                 \centering
                 \epsfig{file=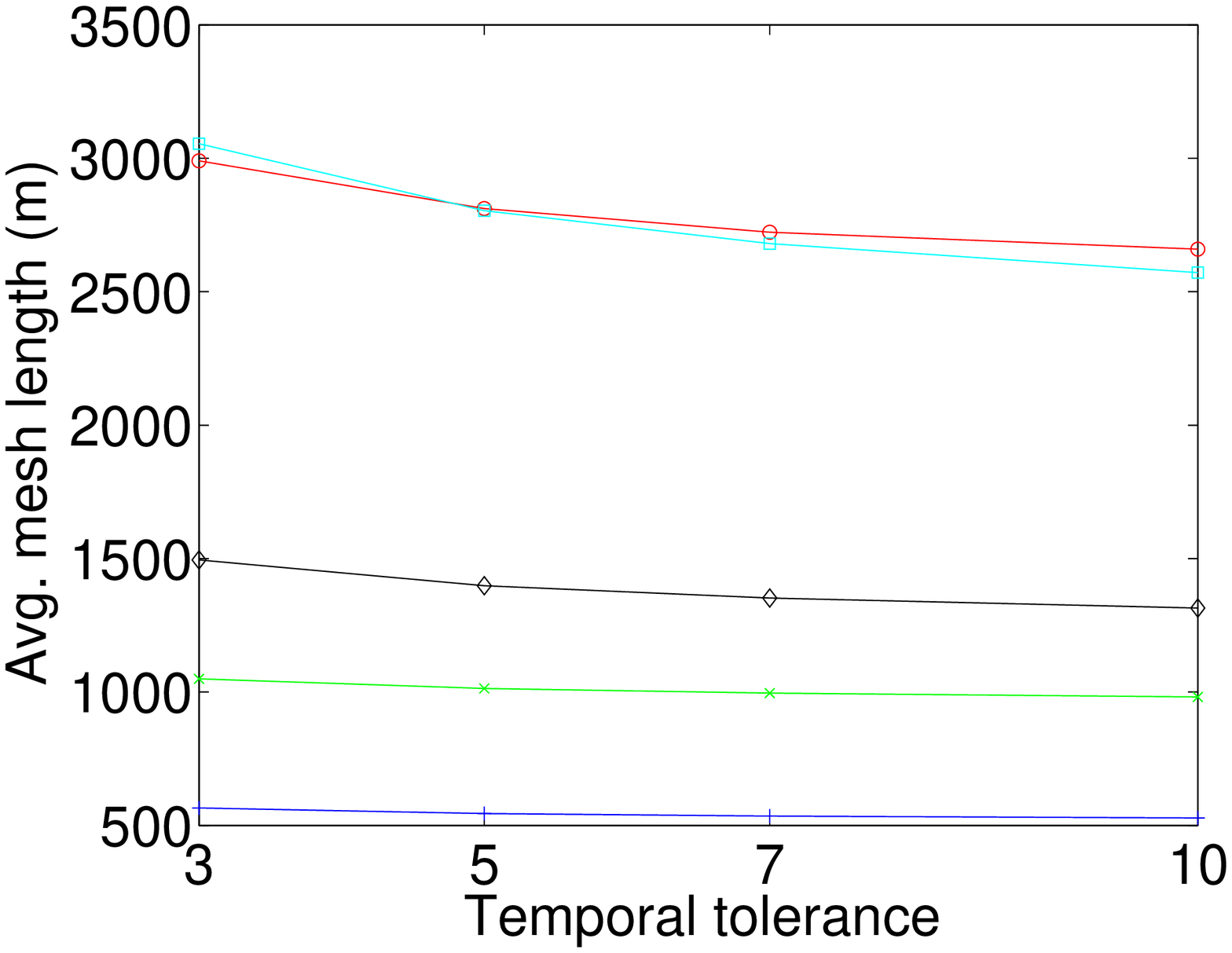, height=1.35in}
                 \setlength{\abovecaptionskip}{-10pt}
                 \caption{P2, k=[2-5]}
%                 \label{fig:util-APD-1}
         \end{subfigure}
         \hfill         
         \begin{subfigure}[b]{0.23\textwidth}
                  \centering
                  \epsfig{file=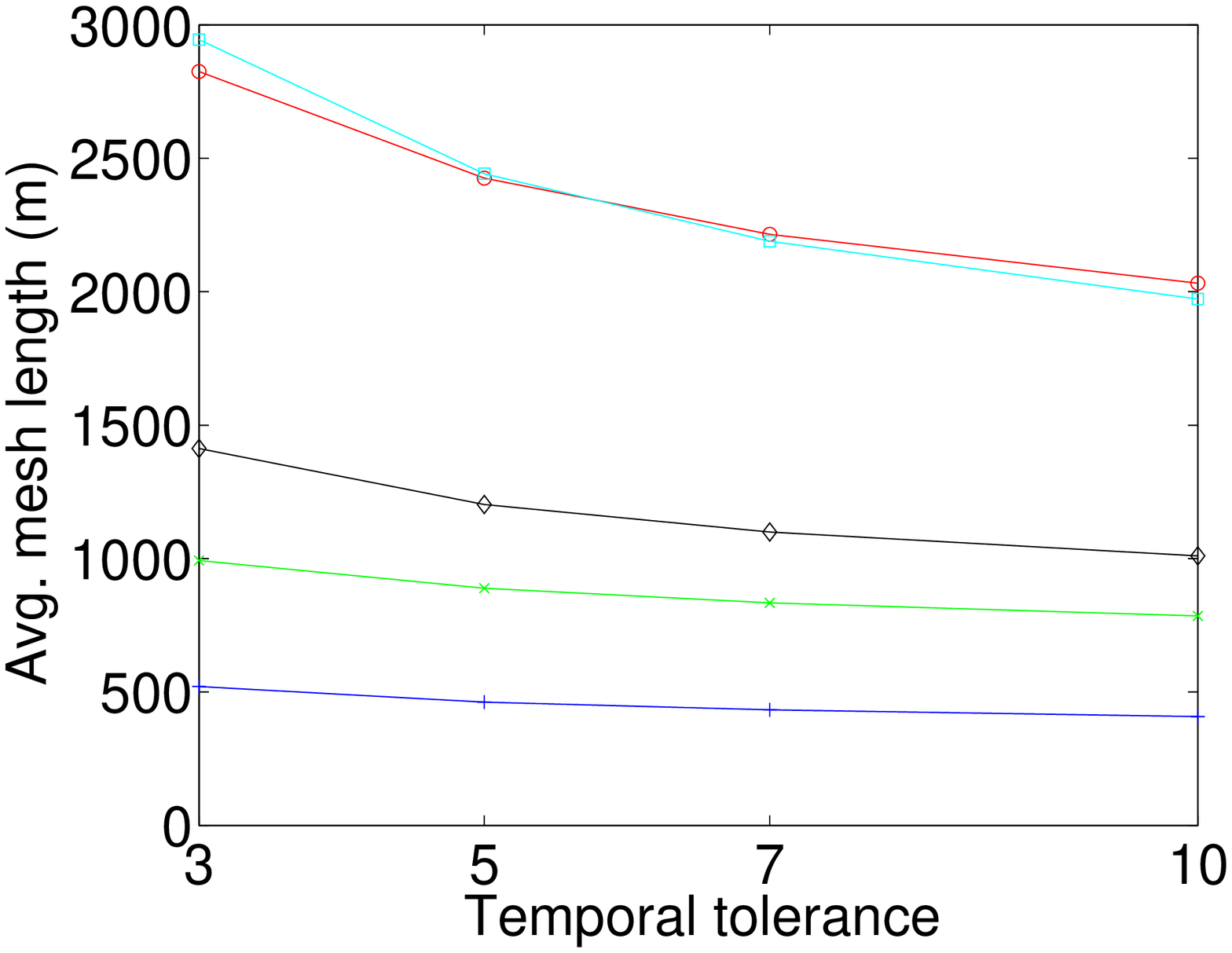, height=1.35in}
                  \setlength{\abovecaptionskip}{-10pt}
                 \caption{P2, k=[2-10]}
%                  \label{fig:util-Dist-1}
         \end{subfigure}        	

     \caption{Average mesh length (the same legends as in Fig. \ref{fig:succ-rate}) }
     \label{fig:avg-ml}
\end{figure*}

\begin{figure*}[t!]
 	\centering
         \begin{subfigure}[b]{0.23\textwidth}
                 \centering
                 \epsfig{file=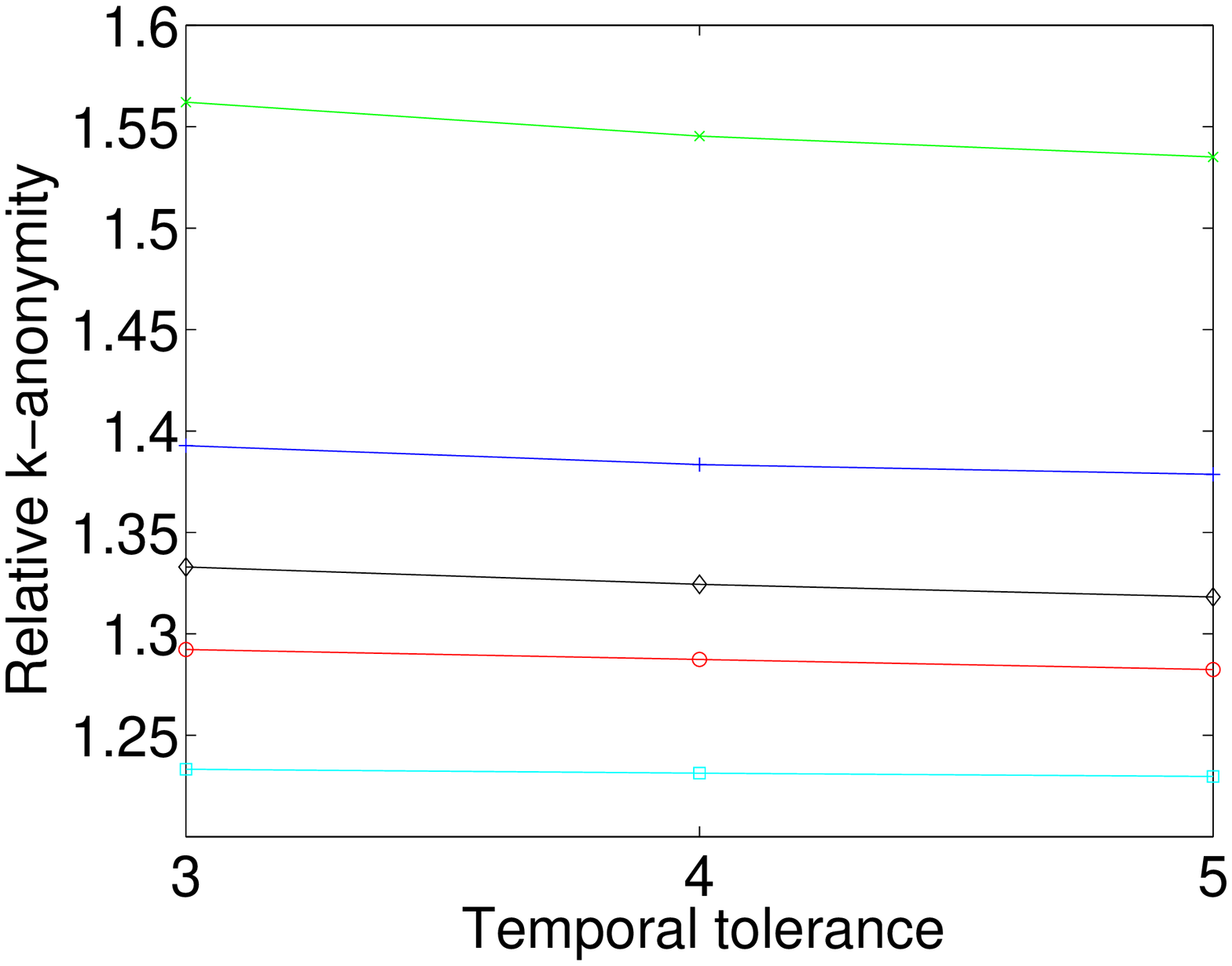, height=1.35in}
                 \setlength{\abovecaptionskip}{-10pt}
                 \caption{P1, k=[2-5]}	
%                 \label{fig:util-PL-1}
         \end{subfigure}
         \hfill
         \begin{subfigure}[b]{0.23\textwidth}
                 \centering
                 \epsfig{file=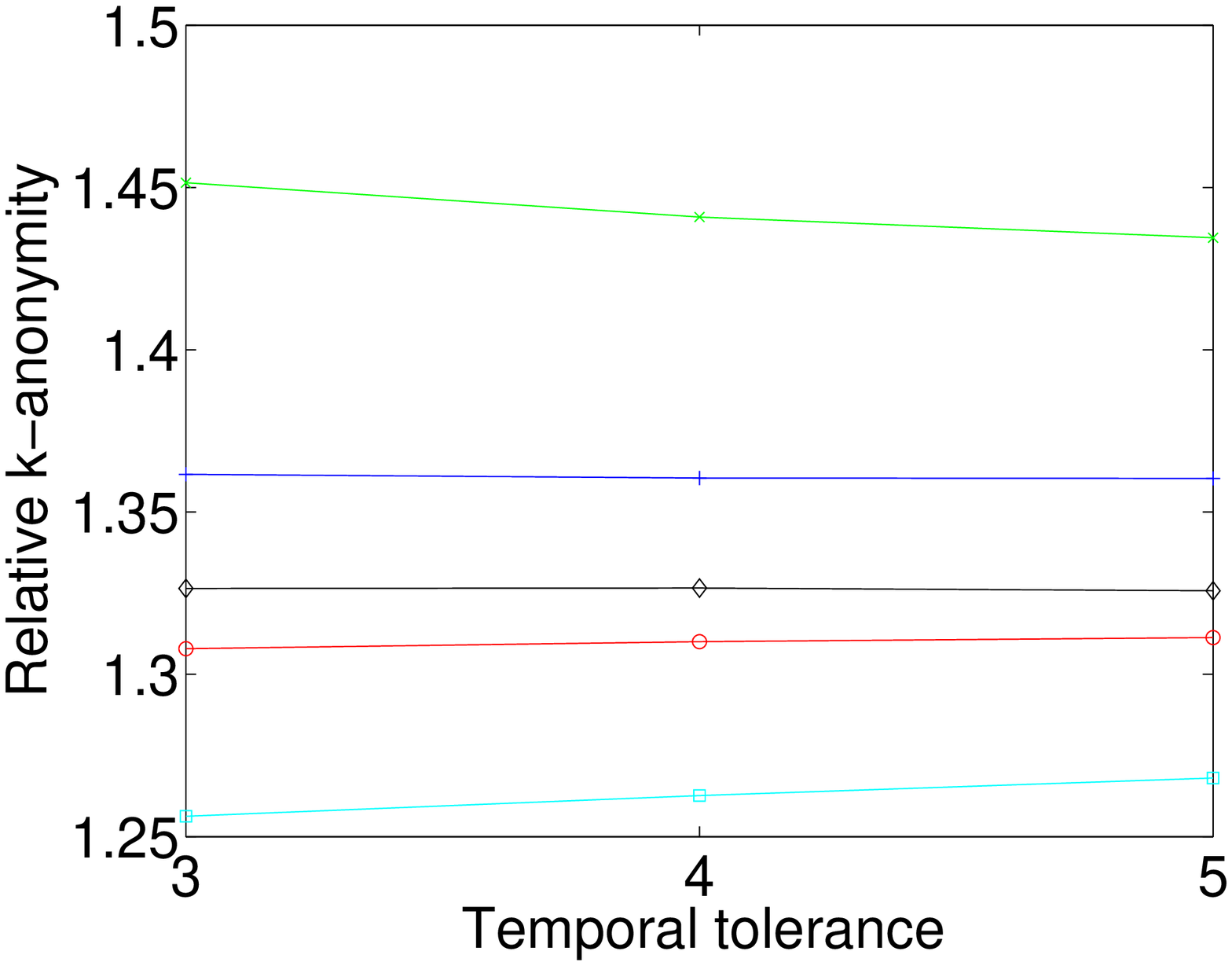, height=1.35in}
                 \setlength{\abovecaptionskip}{-10pt}
                 \caption{P1, k=[2-10]}
%                 \label{fig:util-CC-1}
         \end{subfigure}
         \hfill         
         \begin{subfigure}[b]{0.23\textwidth}
                 \centering
                 \epsfig{file=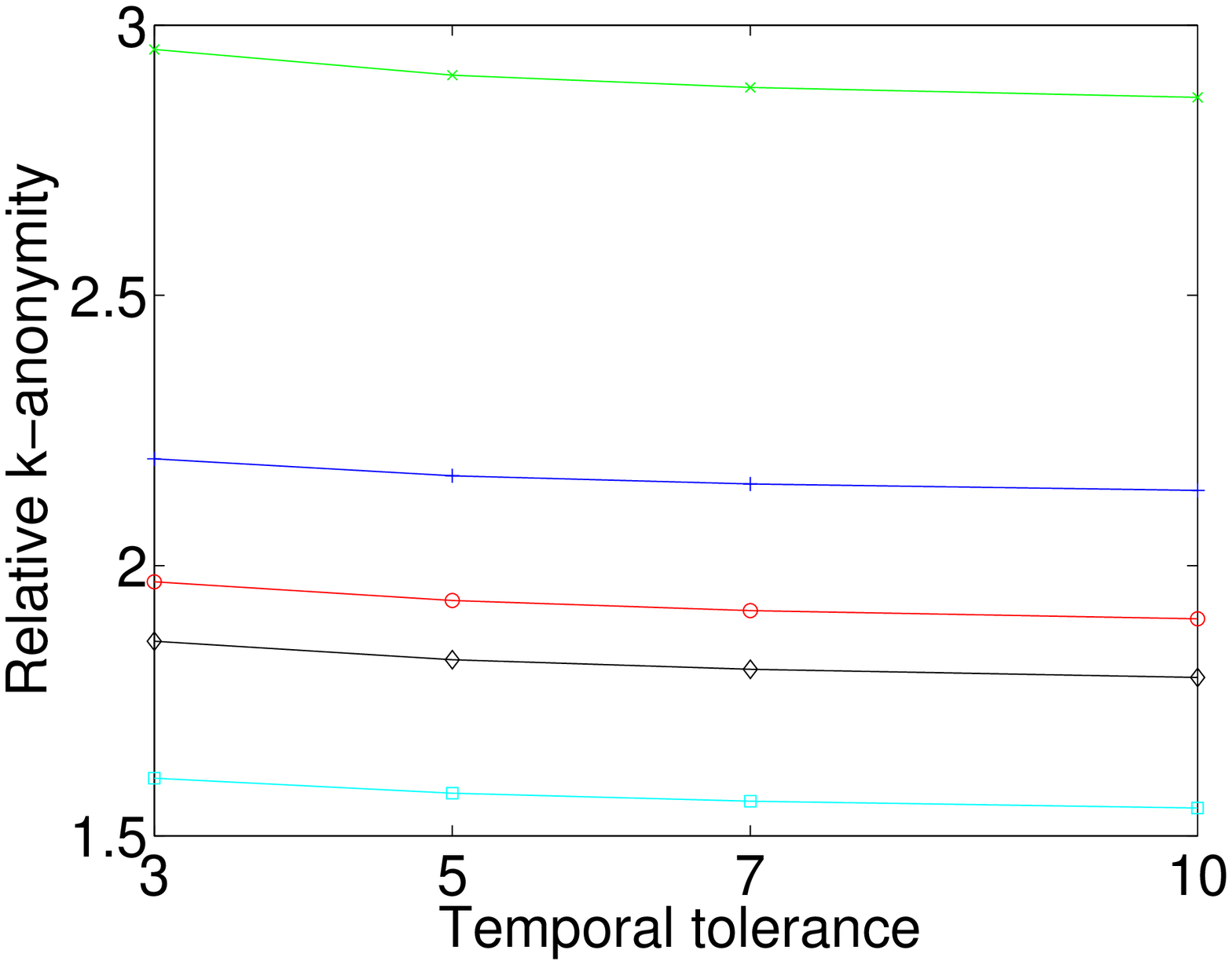, height=1.35in}
                 \setlength{\abovecaptionskip}{-10pt}
                 \caption{P2, k=[2-5]}
%                 \label{fig:util-APD-1}
         \end{subfigure}
         \hfill         
         \begin{subfigure}[b]{0.23\textwidth}
                  \centering
                  \epsfig{file=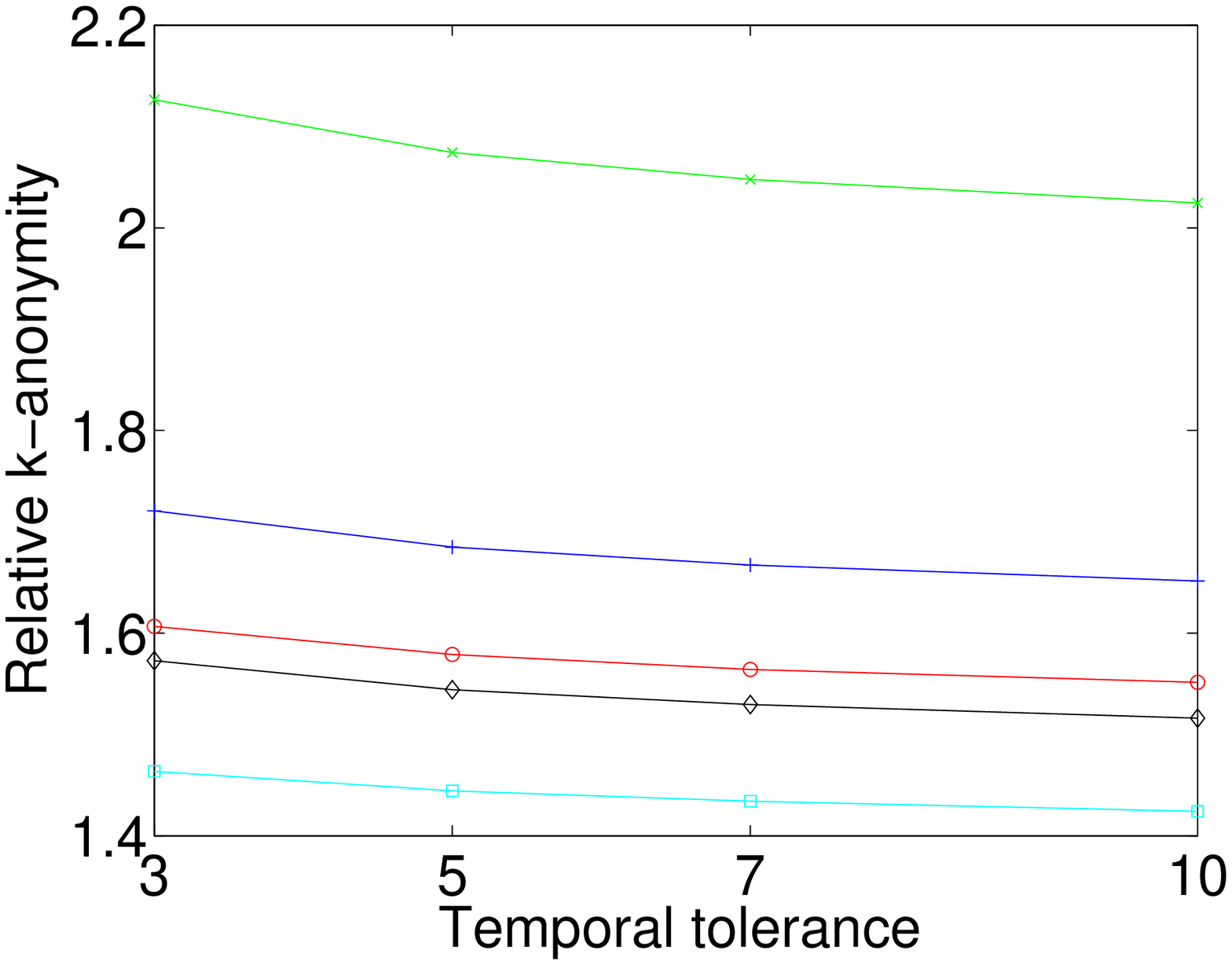, height=1.35in}
                  \setlength{\abovecaptionskip}{-10pt}
                 \caption{P2, k=[2-10]}
%                  \label{fig:util-Dist-1}
         \end{subfigure}        	

     \caption{Relative k-anonymity (the same legends as in Fig. \ref{fig:succ-rate}) }
     \label{fig:rel-k}
\end{figure*}

\begin{figure*}[t!]
 	\centering
         \begin{subfigure}[b]{0.23\textwidth}
                 \centering
                 \epsfig{file=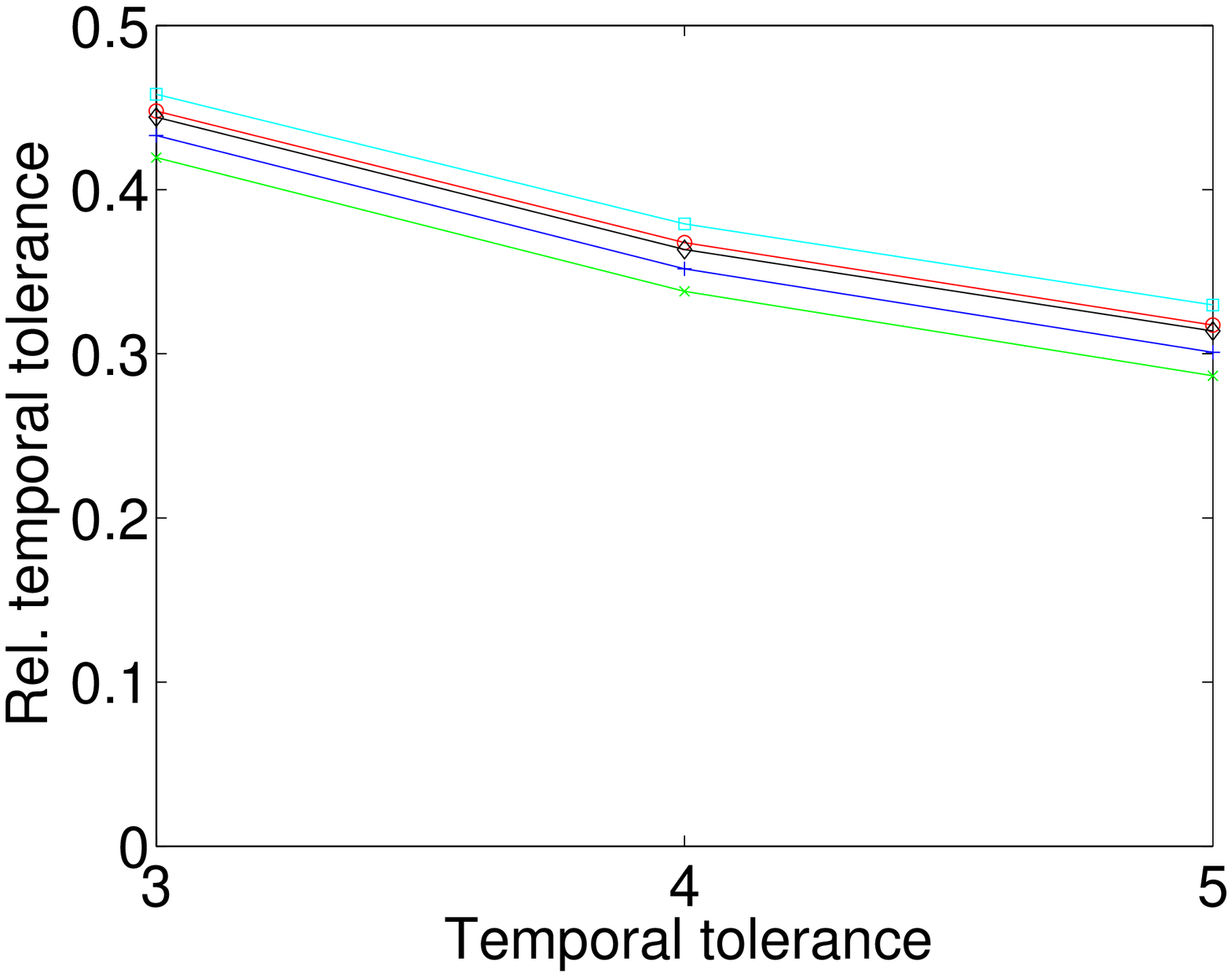, height=1.35in}
                 \setlength{\abovecaptionskip}{-10pt}
                 \caption{P1, k=[2-5]}	
%                 \label{fig:util-PL-1}
         \end{subfigure}
         \hfill
         \begin{subfigure}[b]{0.23\textwidth}
                 \centering
                 \epsfig{file=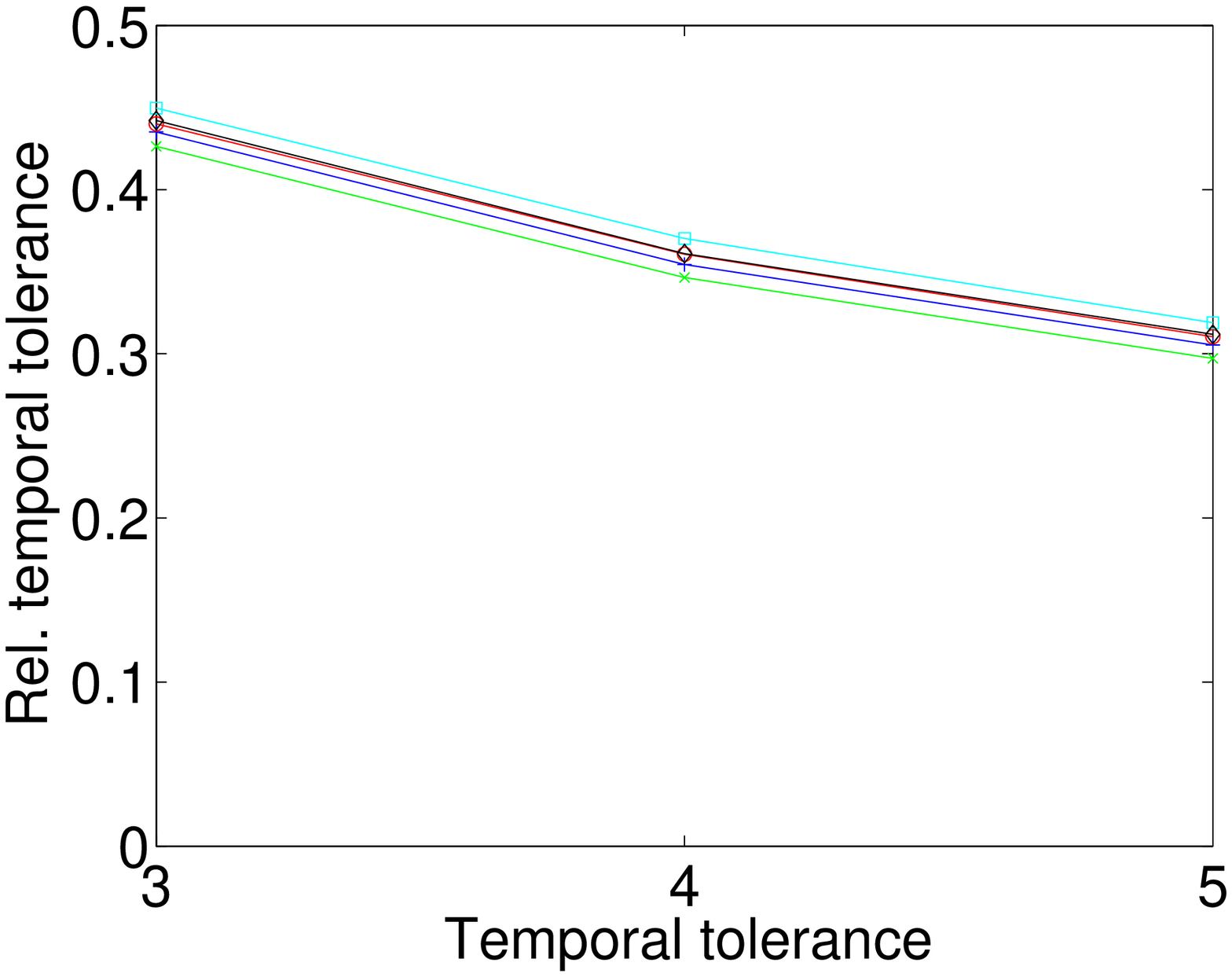, height=1.35in}
                 \setlength{\abovecaptionskip}{-10pt}
                 \caption{P1, k=[2-10]}
%                 \label{fig:util-CC-1}
         \end{subfigure}
         \hfill         
         \begin{subfigure}[b]{0.23\textwidth}
                 \centering
                 \epsfig{file=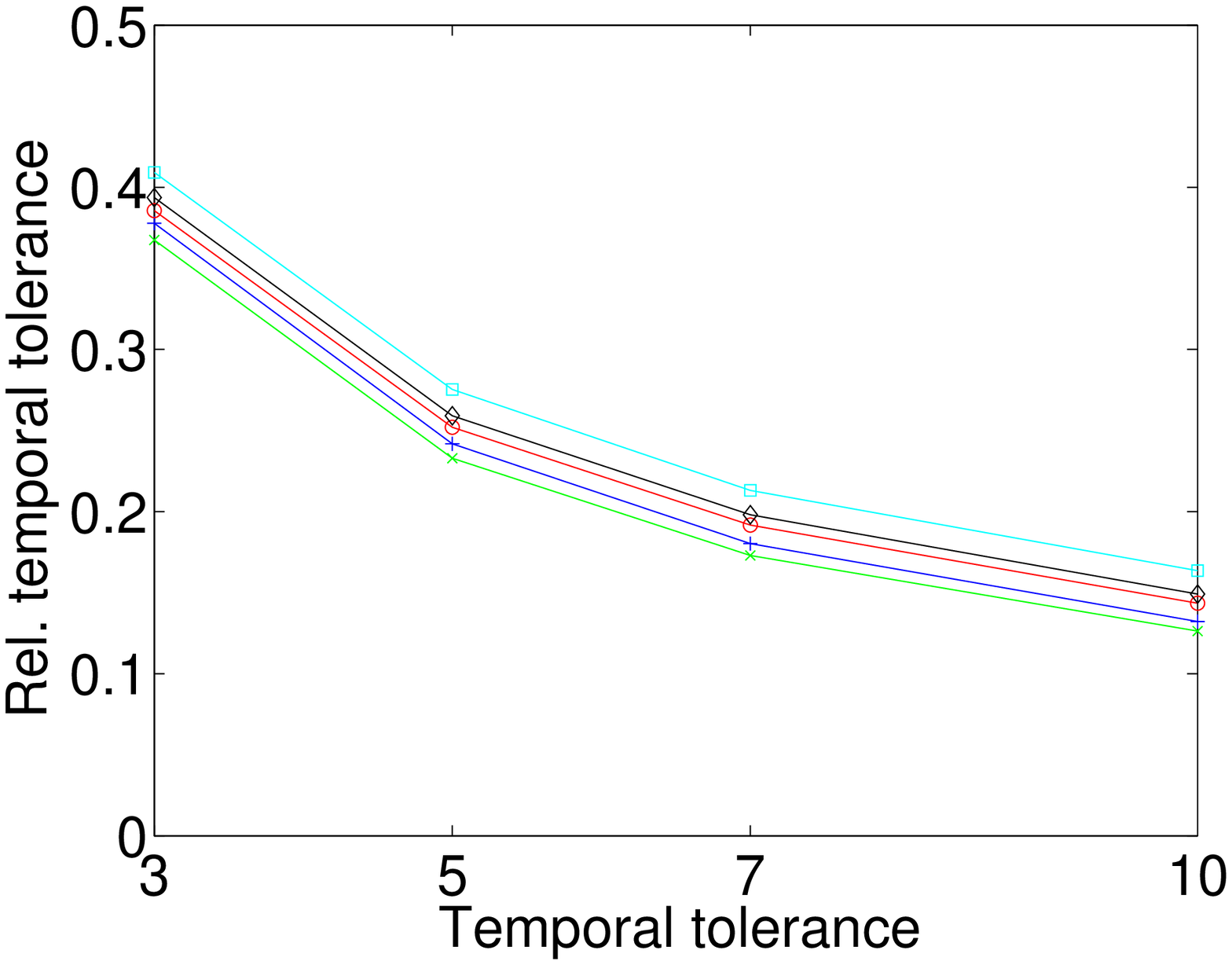, height=1.35in}
                 \setlength{\abovecaptionskip}{-10pt}
                 \caption{P2, k=[2-5]}
%                 \label{fig:util-APD-1}
         \end{subfigure}
         \hfill         
         \begin{subfigure}[b]{0.23\textwidth}
                  \centering
                  \epsfig{file=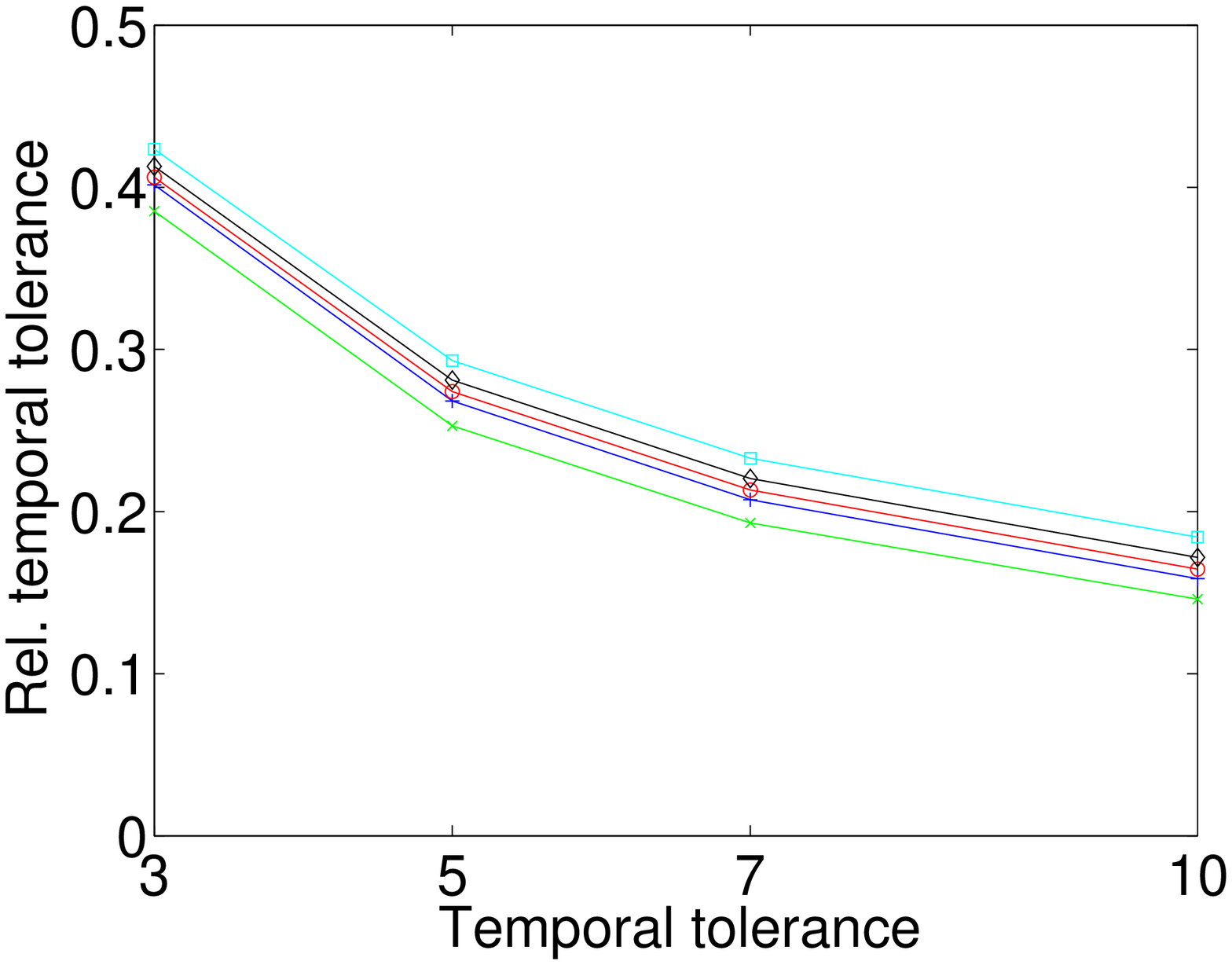, height=1.35in}
                  \setlength{\abovecaptionskip}{-10pt}
                 \caption{P2, k=[2-10]}
%                  \label{fig:util-Dist-1}
         \end{subfigure}        	

     \caption{Relative temporal tolerance (the same legends as in Fig. \ref{fig:succ-rate}) }
     \label{fig:rel-dt}
\end{figure*}

The effectiveness and efficiency of our MeshCloak are reported in Figures \ref{fig:succ-rate}-\ref{fig:rel-dt}. Each metric is evaluated in four combinations of speed profile and k-anonymity at different values of temporal tolerance. For the same speed profile (P1 or P2), we can see that higher k-anonymity levels or smaller temporal tolerances reduce the success rate (Fig. \ref{fig:succ-rate}). This is because smaller temporal tolerances make the WAITING set smaller, so the CG gets smaller and sparser while higher k-anonymity levels cannot be satisfied easily by the size of maximal cliques in the CG. Meanwhile, P2 always gives better success rate than P1 for the same k-anonymity. This is supported by the fact that in P2, the distance constraints are larger, so the CG gets denser, producing larger maximal cliques for cloaking purpose.

The processing time in Fig. \ref{fig:proc-time} highlights the runtime advantage of MeshCloak. Each query is processed in about tenth of millisecond. Our techniques described in Section \ref{sec:algo} are crucial in lowering the time for building the CG. Besides, Tomita algorithm \cite{tomita2006worst} fits well in MeshCloak's batch processing model (one batch per second). As a result, MeshCloak's throughput is much higher than the throughput reported in CliqueCloak and ICliqueCloak in which the average processing time per query is from several to tens of milliseconds.

From the LBS providers' perspective, small to moderate cloaking areas are preferred because large cloaking areas have high impact on the quality of service. Large cloaking areas require longer time to process at LBS servers and heavier data transfer back and forth between the anonymizer and the LBS providers. In our MeshCloak, we use cloaking meshes instead of cloaking areas due to the inherent map-based setting that we advocate. Fig. \ref{fig:avg-ml} displays the average mesh length per query which is about one thousandth of the total street length (cf. Table \ref{tab:map-prop}). Moreover, the average mesh length is quite stable across different speed profiles and k-anonymity levels. Paris-zone1 and Singapore have longer total street length, so are the average mesh length reported in these two maps compared to the three remaining maps.

Finally, relative k-anonymity and relative temporal tolerance are shown in Figures \ref{fig:rel-k} and \ref{fig:rel-dt}. We can observe that the relative k-anonymity is fairly independent of the temporal tolerance. It is always larger than one and rises up for speed profile P2 (cf. \ref{fig:rel-k} c-d). Again, this phenomenon is justified by the denser CG when users move according to P2 profile. Differently, the relative temporal tolerance decreases considerably for larger temporal tolerances. By multiplying the \textit{rel.dt} with the temporal tolerance, we can see that most of the queries are successfully cloaked within two seconds since the time they reach the anonymizer. This fact again guarantees the better service quality because the users would experience only a short waiting time for the LBS results.

\subsubsection{Batch vs. Sequential Processing}
In this section, we compare batch versus sequential processing in MeshCloak. Sequential processing means that for each newly coming query, the anonymizer checks the query's distance constraints and updates the CG. The list of maximal cliques in the CG is incrementally maintained as in ICliqueCloak \cite{pan2012protecting}. However,  we apply \textit{node-based incremental} which is several times faster than the edge-based incremental in ICliqueCloak. Table \ref{tab:batch-seq} compares the processing time per query between the two processing models. Clearly, sequential processing is two orders of magnitude slower than batch processing. This is mainly due to the costly set operations in maintaining maximal cliques by the sequential models. Moreover, taking much more time to process incoming queries in the sequential models impacts heavily the success rate because a large portion of queries will expire. This result again explains our choice of batch model (one batch per second) running the blazingly fast Tomita algorithm \cite{tomita2006worst} on the sparse constraint graphs.

\begin{table}
\centering
\caption{Processing time (ms/query): Batch vs. Sequential processing} \label{tab:batch-seq}
\begin{tabular}{|l|p{15mm}|p{15mm}|r|}
\hline
& \textbf{Batch with \newline Tomita} &\textbf{Node-based \newline Sequential} & \textbf{Seq/Batch}\\
\hline
Oldenburg-P1 (dt=3,k=5) & 0.0137 & 8.09 & 590x \\
\hline
Oldenburg-P1 (dt=5,k=5) & 0.0184 & 12.70 & 690x \\
\hline
Hanoi-P1 (dt=3,k=5) & 0.0338 & 10.06 & 298x \\
\hline
Hanoi-P1 (dt=5,k=5) & 0.0410 & 13.32 & 325x \\
\hline
\end{tabular}
\setlength{\abovecaptionskip}{-5pt}
\setlength{\belowcaptionskip}{-5pt}
\end{table}

\section{Conclusion}
\label{sec:conclusion}
The existing free-space cloaking schemes based on map-based moving patterns suffer from the mismatch between cloaking models and user movements. They omit most of real-world movement constraints. Our paper addresses this shortcoming by stating the problem entirely in the map-based setting and proposes novel scheme to solve it. Our key techniques is a fast distance computation between map nodes (via a pre-computed distance matrix along with quadtree search) and an efficient batch processing model. Experiments in various configurations confirm the effectiveness and efficiency of our MeshCloak in terms of success rate, processing time, average mesh length and relative k-anonymity/temporal tolerance. We believe that the map-based setting deserves more attention in future research on location privacy.


\section*{Conflicts of Interest}
The author(s) declare(s) that there is no conflict of interest regarding the publication of this paper.


\bibliographystyle{abbrv}
\bibliography{meshcloak}  % mesh-cloak.bib is the name of the Bibliography in this case

% This is a hand-made bibliography. If you want to use a BibTeX file, you're on your own ;-)

\end{document}